\documentclass[prd,amsmath,amssymb,superscriptaddress,preprintnumbers,twocolumn,nofootinbib,10pt]{revtex4-1}

\pdfoutput=1

\usepackage{graphicx}
\usepackage{dcolumn}
\usepackage{bm}
\usepackage{amssymb}
\usepackage{latexsym}
\usepackage{booktabs}
\usepackage{amsmath}
\usepackage{multirow}
\usepackage{url}

\usepackage{float}
\usepackage[colorlinks=true, linkcolor=red, citecolor=blue]{hyperref}

\usepackage[normalem]{ulem}
\usepackage{color}
\usepackage{array}
\usepackage{enumerate}

\begin{document}

\title{Prospects for searching for sterile neutrinos with gravitational wave and $\gamma$-ray burst joint observations}

\author{Lu Feng}\thanks{These authors contributed equally to this paper.}
\affiliation{College of Physical Science and Technology, Shenyang Normal University, Shenyang 110034, China}
\affiliation{Liaoning Key Laboratory of Cosmology and Astrophysics, College of Sciences, Northeastern University, Shenyang 110819, China}
\author{Tao Han}\thanks{These authors contributed equally to this paper.}
\affiliation{Liaoning Key Laboratory of Cosmology and Astrophysics, College of Sciences, Northeastern University, Shenyang 110819, China}
\author{Jing-Fei Zhang}
\affiliation{Liaoning Key Laboratory of Cosmology and Astrophysics, College of Sciences, Northeastern University, Shenyang 110819, China}
\author{Xin Zhang}\thanks{Corresponding author}
\email{zhangxin@mail.neu.edu.cn}
\affiliation{Liaoning Key Laboratory of Cosmology and Astrophysics, College of Sciences, Northeastern University, Shenyang 110819, China}
\affiliation{MOE Key Laboratory of Data Analytics and Optimization for Smart Industry, Northeastern University, Shenyang 110819, China}
\affiliation{National Frontiers Science Center for Industrial Intelligence and Systems Optimization, Northeastern University, Shenyang 110819, China}

\begin{abstract}
Sterile neutrinos can influence the evolution of the universe, and thus cosmological observations can be used to detect them. Future gravitational-wave (GW) observations can precisely measure absolute cosmological distances, helping to break parameter degeneracies generated by traditional cosmological observations. This advancement can lead to much tighter constraints on sterile neutrino parameters. This work provides a preliminary forecast for detecting sterile neutrinos using third-generation GW detectors in combination  with future short $\gamma$-ray burst observations from a THESEUS-like telescope, an approach not previously explored in the literature. Both massless and massive sterile neutrinos are considered within the $\Lambda$CDM cosmology. We find that using GW data can greatly enhance the detection capability for massless sterile neutrinos, reaching 3$\sigma$ level. For massive sterile neutrinos, GW data can also greatly assist in improving the parameter constraints, but it seems that effective detection is still not feasible.

\end{abstract}

\maketitle

\section{Introduction}

On 17 August 2017, the first observation of gravitational waves (GW) from a binary neutron star (BNS) merger~\cite{LIGOScientific:2017vwq}, together with the first joint observation of GW from a BNS merger and its electromagnetic (EM) counterpart~\cite{LIGOScientific:2017ync,TOROS:2017pqe}, marked the beginning of a new era in multi-messenger astronomy and cosmology.
The measurement of the GW signal directly provides information on the absolute luminosity distance to the source, while its redshift can be determined by identifying the EM counterpart of the GW source. This establishes an absolute distance-redshift relation, known as the standard siren method, which is crucial for cosmological studies.
To date, only one bright siren, GW170817, has been identified. This is insufficient to probe cosmological parameters using current standard sirens, necessitating the use of next-generation GW detectors.

In the future, third-generation (3G) ground-based GW detectors, such as the Einstein Telescope (ET)~\cite{ET-web,Punturo:2010zz} in Europe and the Cosmic Explorer (CE)~\cite{CE-web,Evans:2016mbw} in the United States will become operational, with sensitivities improved one order of magnitude over the current detectors, and much more BNS merger events  will be observed at much deeper redshifts.
Recently, GW standard sirens have been widely discussed in the literature \cite{Cai:2016sby,Cai:2017plb,Liu:2017xef,Cai:2017aea,Berti:2018cxi,Cai:2018rzd,Wang:2018lun,Zhao:2018gwk,Zhang:2018byx,Du:2018tia,He:2019dhl,Yang:2019bpr,Yang:2019vni,Zhang:2019ylr,Zhang:2019ple,Bachega:2019fki,Wang:2019tto,Zhang:2019loq,Li:2019ajo,Zhao:2019gyk,Jin:2020hmc,Wang:2021srv,Qi:2021iic,Jin:2021pcv,Zhu:2021bpp,deSouza:2021xtg,Wang:2022oou,Wu:2022dgy,Jin:2022qnj,Jin:2022tdf,Hou:2022rvk,Song:2022siz,Jin:2023zhi,Jin:2023sfc,Jin:2023tou,Han:2023exn,Li:2023gtu,Dong:2024bvw,Feng:2024lzh,Bian:2021ini}. It has been found that future observations of GW standard sirens from the ET and CE will play a crucial role in the estimation of cosmological parameters~\cite{Wang:2018lun,Zhang:2018byx,Zhang:2019ple,Zhang:2019loq,Li:2019ajo,Jin:2020hmc,Jin:2022tdf,Hou:2022rvk,Han:2023exn,Feng:2024lzh}. In particular, the GW standard sirens can break the parameter degeneracies generated by the current EM cosmological observations, thereby improving constraints on neutrino mass; see e.g., Refs. \cite{Wang:2018lun,Jin:2022tdf}. 

A recent forecast~\cite{Feng:2024lzh} demonstrated that joint observations of BNS by 3G GW detectors and short $\gamma$-ray burst (GRB) observations by missions similar to the THESEUS satellite project can improve the constraints on the total active neutrino mass. Therefore, it is crucial to investigate how the combined GW-GRB observations would affect constraints on the sterile neutrino parameters. 

The existence of light sterile neutrinos has been suggested by anomalies in short-baseline (SBL) neutrino experiments~\cite{LSND:2001aii,Giunti:2010zu,Akbar:2011qw,Conrad:2012qt,MiniBooNE:2012maf,Giunti:2012tn,Giunti:2012bc,Kopp:2013vaa,Giunti:2013aea,Gariazzo:2013gua}. To explain the SBL neutrino oscillation data, sterile neutrinos with eV-scale masses are required~\cite{Abazajian:2012ys,Hannestad:2012ky,Conrad:2013mka}.
Cosmological observations play a crucial role in constraining the mass of active neutrinos (see e.g., Refs.~\cite{Hu:1997mj,Reid:2009nq,Li:2012vn,Wang:2012vh,Hamann:2012fe,Li:2012spm,Riemer-Sorensen:2013jsa,Giusarma:2013pmn,Cahn:2013taa,Lesgourgues:2014zoa,Zhang:2014nta,Zhou:2014fva,Costanzi:2014tna,Palanque-Delabrouille:2014jca,Zhang:2015rha,Qian:2015waa,Patterson:2015xja,Allison:2015qca,Geng:2015haa,Chen:2015oga,Zhang:2015uhk,Huang:2015wrx,Chen:2016eyp,Moresco:2016nqq,Lu:2016hsd,Hada:2016dje,Wang:2016tsz,Kumar:2016zpg,Zhao:2016ecj,Bohringer:2016fcq,Xu:2016ddc,Vagnozzi:2017ovm,Guo:2017hea,Zhang:2017rbg,Chen:2017ayg,Yang:2017amu,Koksbang:2017rux,Li:2017iur,Wang:2017htc,Zhao:2017jma,Boyle:2017lzt,Vagnozzi:2018jhn,Guo:2018gyo,RoyChoudhury:2018gay,Feng:2019mym,Zhang:2019ipd,Li:2020gtk,Zhang:2020mox}).
Since sterile neutrinos have implications for the evolution of the universe, cosmology can provide an independent test for their existence. For related works on sterile neutrinos, see e.g., Refs.~\cite{deHolanda:2010am,Palazzo:2013me,Hamann:2013iba,Wyman:2013lza,Battye:2013xqa,Dvorkin:2014lea,Archidiacono:2014apa,Ko:2014bka,Li:2014dja,Zhang:2014dxk,Archidiacono:2014nda,Bergstrom:2014fqa,DayaBay:2014fct,Zhang:2014ifa,Zhang:2014lfa,Li:2015poa,Feng:2017nss,Zhao:2017urm,Feng:2017mfs,Feng:2017usu,Knee:2018rvj,Feng:2019jqa,Feng:2021ipq,DiValentino:2021rjj,Chernikov:2022mdn,Pan:2023frx}.

Currently, one of the most significant challenges in cosmology is the “Hubble tension”~\cite{Verde:2019ivm}, which refers to the discrepancy between early and late universe observations. 
In the past few years, people often considered models that include light sterile neutrinos to alleviate the Hubble constant crisis. This is because when using the cosmic microwave background data to constrain cosmological parameters, the effective number of neutrino species ($N_{\rm eff}$) is positively correlated with the Hubble constant; if $N_{\rm eff}$ is larger, the derived Hubble constant will also be larger. See e.g., Refs.~\cite{Zhang:2014dxk,Feng:2017nss,Zhao:2017urm,Feng:2017mfs,Feng:2017usu,Feng:2019jqa,Feng:2021ipq,Pan:2023frx}, for related studies. However, in recent years, the results of cosmological observation fits have shown that the effect of using this method to alleviate the Hubble crisis is no longer significant. Nevertheless, due to the significant correlation between $N_{\rm eff}$ and $H_0$, precise measurements of the Hubble constant using gravitational wave standard sirens are very helpful for determining the parameters of sterile neutrinos.


Additionally, the main advantage of the standard siren method for measuring the Hubble constant is that it avoids relying on the cosmic distance ladder. Therefore, in the future, the GW standard sirens could become a promising cosmological probe, playing a crucial role in measuring cosmological parameters, including those related to sterile neutrinos.

In this paper, we present a forecast for the search for sterile neutrinos using joint GW-GRB observations. 
The primary aim of this work is to investigate the impact of future GW standard siren observations on the constraints of sterile neutrino parameters.

This work is organized as follows. In Sec.~\ref{sec2}, we introduce the methodology used in this work. In Sec.~\ref{sec3}, we give the constraint results and make some relevant discussions. The conclusion is given in Sec.~\ref{sec4}. 

\section{Methodology}\label{sec2}

\subsection{Gravitational wave simulation}\label{sec2.1}
In this subsection, we introduce the method of simulating the joint GW standard sirens and GRB events. We consider the THESEUS-like GRB detector in synergy with the 3G GW observation. We use the simulation method as prescribed in Refs.~\cite{Han:2023exn,Feng:2024lzh}. Here, we provide only a brief overview.

The BNS merger rate with redshift in the observer frame is~\cite{Vitale:2018yhm,Yang:2021qge,Belgacem:2019tbw}
\begin{equation}
R_{\rm m}(z)=\frac{\mathcal{R}_{\rm m}(z)}{1+z} \frac{{\rm d}V(z)}{{\rm d}z},
\label{eq:1}
\end{equation}
where $dV(z)/dz$ is the comoving volume element, the factor $(1+z)^{-1}$ converts the merger rate in the source frame to the observer frame, and $\mathcal{R}_{\rm m}(z)$ is the BNS merger rate in the source frame, expressed as
\begin{equation}
	\mathcal{R}_{\rm m}(z)=\int_{t_{\rm min}}^{t_{\rm max}} \mathcal{R}_{\rm f}[t(z)-t_{\rm d}] P(t_{\rm d}){\rm d}t_{\rm d},
	\label{eq:2}
\end{equation}
which is commonly used in the literature \cite{Belgacem:2019tbw,Chen:2018rzo,Du:2021fmb,deSouza:2019ype,Regimbau:2016ike,Yang:2021qge,Belgacem:2018lbp,Safarzadeh:2019pis,Song:2019ddw,Wanderman:2014eza,Yu:2021nvx}\footnote{Here, we take into account the time dilation factor $(1+z)^{-1}$ in Eq.~(\ref{eq:1}), which differs from the expression presented in Ref.~\cite{Regimbau:2014nxa}. In this paper, we adopt the commonly used formula in the literature, which is somewhat different from that given in Ref.~\cite{Regimbau:2014nxa}.}.
Here, $t_{\rm d}$ is the delay time between the formation of BNS system and merger, $t_{\rm min}=20$ Myr is the minimum delay time, $t_{\rm max}=t_{\rm H}$ is the maximum delay time, $t(z)$ is the age of the universe at the time of merger, $\mathcal{R}_{\rm f}$ is the cosmic star formation rate in the source frame for which we adopt the Madau-Dickinson model~\cite{Madau:2014bja}, $P(t_{\rm d})$ is the time delay distribution of the $t_{\rm d}$, and we adopt the exponential time delay model~\cite{Vitale:2018yhm}, which is given by
\begin{equation}
	P(t_{\rm d})=\frac{1}{\tau}{\rm exp}(-t_{\rm d}/\tau),
\end{equation}
with an e-fold time of $\tau=0.1$ Gyr for $t_{\rm d}>t_{\rm min}$.

In our calculations, for BNS mergers, we consider the local comoving merger rate to be $\mathcal{R}_{\rm m}(z=0)=920~\rm Gpc^{-3}~yr^{-1}$, which is the estimated median from the O1 LIGO and the O2 LIGO/Virgo observation run~\cite{Eichhorn:2018phj} and is also consistent with the O3 observation run~\cite{KAGRA:2021duu}.
We simulate a catalog of BNS mergers for 10 years observation.
For each source, the location $(\theta,\phi)$, the polarization angle $\psi$, the cosine of the inclination angle $\iota$, and the coalescence phase $\psi_{\rm c}$ are drawn from uniform distributions. Currently, there are multiple candidate models for the neutron star (NS) mass distribution. However, different mass distributions of NSs have less impact on the cosmological analysis~\cite{Han:2023exn}. For simplicity, we employ a Gaussian mass distribution. This distribution has a mean of $1.33~M_{\odot}$ for the NS mass and a standard deviation of $0.09~M_{\odot}$, where $M_{\odot}$ represents the solar mass~\cite{LIGOScientific:2018mvr,Ozel:2016oaf}.

Under the stationary phase approximation~\cite{Zhang:2017srh}, the Fourier transform of the frequency-domain GW waveform for a detector network (with $N$ detectors) is given by~\cite{Zhao:2017cbb,Wen:2010cr}
\begin{equation}
\tilde{\boldsymbol{h}}(f)={\rm e}^{{\rm -i}\boldsymbol\Phi}\boldsymbol h(f),
\end{equation}
with the $\boldsymbol h$($f$) is given by
\begin{equation}
\boldsymbol h(f)=\Big[\frac{h_1(f)}{\sqrt{S_{\rm {n},1}(f)}}, \frac{h_2(f)}{\sqrt{S_{\rm {n},2}(f)}}, \ldots,\frac{h_N(f)}{\sqrt{S_{{\rm n},N}(f)}}\Big ]^{\rm T},
\end{equation}
where $\boldsymbol\Phi$ is the $N\times N$ diagonal matrix with $\Phi_{ij}=2\pi f\delta_{ij}(\boldsymbol{n\cdot r}_k)$, $\boldsymbol n$ is the propagation direction of GW, and $\boldsymbol r_k$ is the location of the $k$-th detector. Here $S_{{\rm n},k}(f)$ is the one-side noise power spectral density of the $ k$-th detector, The Fourier transform of the GW waveform of $ k$-th detector is given by
\begin{align}
	h_k(f)=&\mathcal A_k f^{-7/6}{\rm exp}
	\{{\rm i}[2\pi f t_{\rm c}-\pi/4-2\psi_c+2\Psi(f/2)]\nonumber\\ &-\varphi_{k,(2,0)})\},
\end{align}
where the Fourier amplitude can be written as
\begin{align}
	\mathcal A_k=&\frac{1}{d_{\rm L}}\sqrt{(F_{+,k}(1+\cos^{2}\iota))^{2}+(2F_{\times,k}\cos\iota)^{2}}\nonumber\\ &\times\sqrt{5\pi/96}\pi^{-7/6}\mathcal M^{5/6}_{\rm chirp}.
\end{align}
Here, the detailed forms of $\Psi(f/2)$ and $\varphi_{k,(2,0)}$ can be found in Refs.~\cite{Cutler:1992tc,Zhao:2017cbb}, $d_{\rm L}$ is the luminosity distance of the GW source, $\mathcal M_{\rm chirp}=(1+z)\eta^{3/5}M$ is the observed chirp mass, $\eta=m_1 m_2/M^2$ is the symmetric mass ratio, and $M=m_1+m_2$ is the total mass of binary system with component masses $m_1$ and $m_2$, $F_{+, k}$ and $F_{\times,k}$ are the antenna response functions of the $k$-th GW detector, we adopt the GW waveform in the frequency domain, in which the time $t$ is replaced by $t_{\rm f}=t_{\rm c}-(5 / 256) \mathcal{M}_{\rm chirp}^{-5 / 3}(\pi f)^{-8 / 3}$~\cite{Cutler:1992tc,Zhao:2017cbb}, where $t_{\rm c}$ is the coalescence time.

In this work, we consider the waveform in the inspiralling stage for the non-spinning BNS system. Here we adopt the restricted Post-Newtonian approximation and calculate the waveform to the 3.5 PN order~\cite{Cutler:1992tc,Sathyaprakash:2009xs}. 

After simulating the GW catalog, we need to calculate the signal-to-noise ratio (SNR) for each GW event. The SNR for the detection network of $N$ independent interferometers can be calculated by
\begin{equation}
\rho=(\tilde{\boldsymbol h}|\tilde{\boldsymbol h})^{1/2}.
\end{equation}
The inner product is defined as
\begin{equation}
(\boldsymbol a|\boldsymbol b)=2\int_{f_{\rm lower}}^{f_{\rm upper}}\{\boldsymbol a(f)\boldsymbol b^*(f)+\boldsymbol a^*(f)\boldsymbol b(f)\}{\rm d}f,
\end{equation}
where $*$ represents conjugate transpose, $\boldsymbol a$ and $\boldsymbol b$ are column matrices of the same dimension, the lower cutoff frequency is set to $f_{\rm lower}=1$ Hz for ET and $f_{\rm lower}=5$ Hz for CE, and $f_{\rm upper}=2/(6^{3/2}2\pi M_{\rm obs})$ is the frequency at the last stable orbit with $M_{\rm obs}=(m_1+m_2)(1+z)$. In this work, we adopt the SNR threshold to be 12 in our simulation.

For the short GRB model, we adopt the model of Gaussian structured jet profile based on the GW170817/GRB170817A~\cite{Howell:2018nhu} observation,
\begin{equation}
L_{\rm iso}(\theta_{\rm v})=L_{\rm on}\exp\left(-\frac{\theta^2_{\rm v}}{2\theta^2_{\rm c}} \right),
\label{eq:jet}
\end{equation}
where $\theta_{\rm v}$ is the viewing angle, $L_{\rm iso}(\theta_{\rm v})$ is the isotropically equivalent luminosity of short GRB observed at different $\theta_{\rm v}$, $L_{\rm on}=L_{\rm iso}(0)$ is the on-axis isotropic luminosity, $\theta_{\rm c}=4.7^{\circ}$ is the characteristic angle of the core, and the direction of the jet is assumed to aligne with the binary orbital angular momentum, namely $\iota=\theta_{\rm v}$.

For the distribution of the short GRB, we assume the empirical broken-power-law luminosity function\footnote{There are several realistic candidates for the luminosity function of short GRB in recent studies; see, e.g., Ref.~\cite{Tan:2020vtc} for more detailed discussions.}
\begin{equation}
	\Phi(L)\propto
	\begin{cases}
		(L/L_*)^{\alpha}, & L<L_*, \\
		(L/L_*)^{\beta}, & L\ge L_*,
	\end{cases}
	\label{eq:distribution}
\end{equation}
which is commonly used in the literature~\cite{Belgacem:2019tbw,Wanderman:2014eza,Yang:2021qge,Jin:2023tou,Feng:2024lzh,Han:2023exn,Hou:2022rvk,Regimbau:2014nxa}. Here $L$ is the isotropic rest frame luminosity in the $1-10000$ keV energy range, $L_{*}$ is the characteristic luminosity that separates the low and high end of the luminosity function, and the slopes describing these regimes are given by $\alpha$ and $\beta$, respectively.
Following Ref.~\cite{Wanderman:2014eza}, we adopt $L_{*}=2\times10^{52}$ erg sec$^{-1}$, $\alpha=-1.95$, and $\beta=-3$.
We assume a standard low end cutoff in luminosity of $L_{\rm min} = 10^{49}$ erg sec$^{-1}$, and we also term the on-axis isotropic luminosity $L_{\rm on}$ as the peak luminosity $L$~\cite{Belgacem:2019tbw,Tan:2020vtc,Yang:2021qge,Han:2023exn}.
For the THESEUS mission~\cite{Stratta:2018ldl}, a GRB detection is recorded if the value of observed flux is greater than the flux threshold $P_{\rm T}=0.2~\rm ph~s^{-1}~cm^{-2}$ in the 50-300 keV band.
For the GRB detection, we assume a duty cycle of 80\% and a sky coverage fraction of 0.5.
From the GW catalogue which has passed the threshold 12, we can select the GW-GRB events according to the probability distribution $\Phi(L){\rm d}L$.

For a network with $N$ independent interferometers, the Fisher information matrix is given by
\begin{equation}
F_{ij}=\left(\frac{\partial \tilde{\boldsymbol{h}}}{\partial \theta_i}\Bigg |\frac{\partial \tilde{\boldsymbol{h}}}{\partial \theta_j}\right),
\end{equation}
where $\theta_i$ denotes nine GW parameters ($d_{\rm L}$, $\mathcal{M}_{\rm chirp}$, $\eta$, $\theta$, $\phi$, $\iota$, $t_{\rm c}$, $\psi_{\rm c}$, $\psi$) for a GW event. 

For the total uncertainty of the luminosity distance $d_{\rm L}$, we first consider the instrumental error $\sigma_{d_{\rm L}}^{\rm inst}$. The covariance matrix is equal to the inverse of the Fisher information matrix, thus the instrumental error of GW parameter $\theta_i$ is
\begin{equation}
\Delta \theta_i =\sqrt{(F^{-1})_{ii}},
\end{equation}
where $F_{ij}$ is the total Fisher information matrix for the network of $N$ interferometers.

In addition, the weak-lensing error $\sigma_{d_{\rm L}}^{\rm lens}$ and the peculiar velocity error $\sigma_{d_{\rm L}}^{\rm pv}$ are also considered.
The error caused by weak lensing is adopted from Refs.~\cite{Speri:2020hwc,Hirata:2010ba},
\begin{align}
\sigma_{d_{\rm L}}^{\rm lens}(z)=&\left[1-\frac{0.3}{\pi/2} \arctan(z/0.073)\right]\times d_{\rm L}(z)\nonumber\\ &\times 0.066\left [\frac{1-(1+z)^{-0.25}}{0.25}\right ]^{1.8}.\label{lens}
\end{align}

For the error caused by the peculiar velocity of the GW source is given by~\cite{Kocsis:2005vv}
\begin{equation}
	\sigma_{d_{\rm L}}^{\rm pv}(z)=d_{\rm L}(z)\times \left [ 1+ \frac{c(1+z)^2}{H(z)d_{\rm L}(z)}\right ]\frac{\sqrt{\langle v^2\rangle}}{c},\label{pv}
\end{equation}
where $c$ is the speed of light in vacuum, $H(z)$ is the Hubble parameter, and $\sqrt{\langle v^2\rangle}$ is the peculiar velocity of the GW source with respect to the Hubble flow is roughly set to $\sqrt{\langle v^2\rangle}=500\ {\rm km\ s^{-1}}$.

Hence, the total error of $d_{\rm L}$ can be written as
\begin{align}
\sigma_{d_{\rm L}}&~~=\sqrt{(\sigma_{d_{\rm L}}^{\rm inst})^2+(\sigma_{d_{\rm L}}^{\rm lens})^2+(\sigma_{d_{\rm L}}^{\rm pv})^2}.\label{total}
\end{align}

\subsection{Other cosmological observations}\label{sec2.2}
For comparison, we also employ three current EM cosmological observations, i.e., the cosmic microwave background (CMB) data, the baryon acoustic oscillation (BAO) data, and the type Ia supernova (SN) data. The details of these data are listed as follows.

{\it The CMB data}: the CMB likelihood including the TT, TE, EE spectra at $l\geq 30$, the low-$l$ temperature commander likelihood, and the low-$l$ SimAll EE likelihood, from the Planck 2018 data release~\cite{Planck:2018vyg}.

{\it The BAO data}: the measurements from 6dFGs at $z_{\rm eff}=0.106$~\cite{Beutler:2011hx}, the SDSS-MGS at $z_{\rm eff}=0.15$~\cite{Ross:2014qpa}, and BOSS-DR12 at $z_{\rm eff}=0.38$ , $z_{\rm eff}=0.51$, and $z_{\rm eff}=0.61$~\cite{BOSS:2016wmc}.

{\it The SN data}: the Pantheon sample comprised of 1048 data points from the Pantheon complation~\cite{Pan-STARRS1:2017jku}.

\subsection{Methods of constraining cosmological parameters}\label{sec2.3}
In this paper, we will consider both cases of massless and massive sterile neutrinos in the framework of standard model ($\Lambda$CDM) of cosmology. For the $\Lambda$CDM model, the base parameter set (including six free parameters) is
$${\bf P}=\{\omega_b,~\omega_c,~100\theta_{\rm MC},~\tau,~\ln (10^{10}A_s),~n_s\},$$
where $\omega_b\equiv \Omega_b h^2$ and $\omega_c\equiv \Omega_c h^2$ are the physical densities of baryon and cold dark matter, respectively, $\theta_{\rm MC}$ is the ratio (multiplied by 100) between the sound horizon and the angular diameter distance at the time of last-scattering, $\tau$ is the optical depth to the reionization, $A_s$ is the amplitude of the primordial curvature perturbation, and $n_s$ is the scalar spectral index. 

When we consider massless sterile neutrinos (as the dark radiation) in the $\Lambda$CDM model, an additional parameter $N_{\rm eff}$ (the effective number of relativistic species) need to be added in the model, and this case is called $\Lambda$CDM+$N_{\rm eff}$ model in this paper.
When the massive sterile neutrinos are considered in the $\Lambda$CDM model, two extra free parameters, the $N_{\rm eff}$ and $m_{\nu,{\rm sterile}}^{\rm eff}$ (the effective sterile neutrino mass) need to be added in the model, and this case is called $\Lambda$CDM+$N_{\rm eff}$+$m_{\nu,{\rm{sterile}}}^{\rm{eff}}$ model in this paper.
Thus, the $\Lambda$CDM+$N_{\rm eff}$ model has seven independent parameters, and the $\Lambda$CDM+$N_{\rm eff}$+$m_{\nu,{\rm{sterile}}}^{\rm{eff}}$ has eight independent parameters. Note that in both the massless and massive sterile neutrino cases the total mass of active neutrinos is fixed at $\sum m_\nu = 0.06$ eV.

For the GW standard siren observation with $N$ data point, the $\chi^2$ function is defined as
\begin{align}
\chi_{\rm GW}^2=\sum\limits_{i=1}^{N}\left[\frac{{d}_{\rm L}^i-d_{\rm L}({z}_i;\vec{\Omega})}{{\sigma}_{d_{\rm L}}^i}\right]^2,
\label{equa:chi2}
\end{align}
where ${z}_i$, ${d}_{\rm L}^i$, and ${\sigma}_{d_{\rm L}}^i$ are the $i$-th GW redshift, luminosity distance, and the measurement error of the luminosity distance, respectively, $\vec{\Omega}$ denotes the set of cosmological parameters.

\begin{figure}[htbp]
	\includegraphics[width=0.9\linewidth,angle=0]{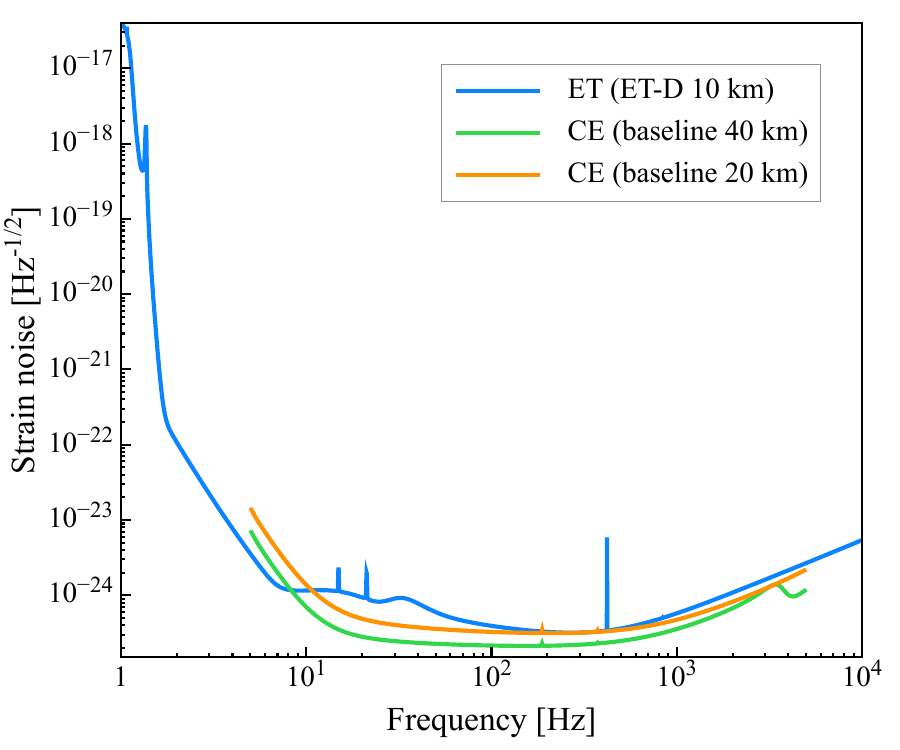}
	\caption{\label{fig1} Sensitivity curves of the 3G GW detectors considered in this work.}
\end{figure}

\begin{figure}[htbp]
	\includegraphics[width=0.9\linewidth,angle=0]{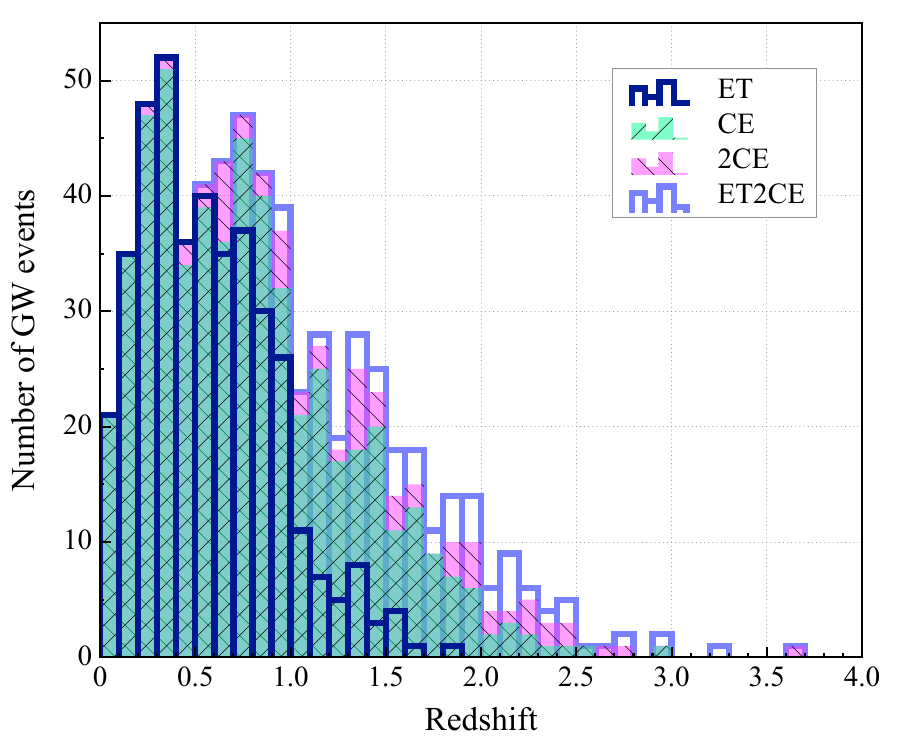}
	\caption{\label{fig2} Redshift distributions of BNS detected by THESEUS in synergy with ET, CE, 2CE, and ET2CE for a 10-year observation.}
\end{figure}

In this work, we present the first forecast for the search for sterile neutrinos using joint GW-GRB observation. We use the public Markov-chain Monte Carlo (MCMC) package CosmoMC~\cite{Lewis:2002ah} to constrain sterile neutrino and other cosmological parameters.
To demonstrate the impact of simulated GW data on constraining sterile neutrino parameters, we will consider all the different cases of 3G GW observations, the single ET, the single CE, the CE-CE network (one CE in the United States with 40-km arm length and another one in Australia with 20-km arm length, abbreviated as 2CE hereafter), and the ET-CE-CE network (one ET detector and two CE-like detectors, abbreviated as ET2CE hereafter) to analysis. 
We utilize the sensitivity curves of ET from Ref.~\cite{ETcurve-web} and for CE from Ref.~\cite{CEcurve-web}, as shown in Fig.~\ref{fig1}. For the GW detector, in view of the high uncertainty of the duty cycle, we only calculate the ideal scenario assuming a 100\% duty cycle for all detectors, as discussed in Ref.~\cite{Zhu:2021ram}. 
The specific parameters characterizing the geometry of GW detector (latitude $\varphi$, longitude $\lambda$, opening angle $\zeta$, and arm bisector angle $\gamma$) are detailed in Table~\ref{tabgw}. The number of GW standard sirens in the subsequent cosmological analysis are shown in Table~\ref{tabsts} and their redshift distributions are shown in Fig.~\ref{fig2}.

\begin{table*}[!htb]
	\caption{The specific coordinate parameters considered in this work.}
	\label{tabgw} 
	\setlength\tabcolsep{20pt}
	\renewcommand{\arraystretch}{1.5}
	\begin{tabular}{ccccc}
		\hline 
		GW detector  & $\varphi\ (\mathrm{deg})$ & $\lambda\ (\mathrm{deg})$ & $\gamma\ (\mathrm{deg})$ & $\zeta\ (\mathrm{deg})$ \\
		\hline
		Einstein Telescope, Europe &  $40.443$ & $9.457$ & $0.000$ & 60 \\
		Cosmic Explorer, USA &  $43.827$ & $-112.825$ & $45.000$ & 90 \\
		Cosmic Explorer, Australia &  $-34.000$ & $145.000$ & $90.000$ & 90 \\
		\hline
	\end{tabular}
\end{table*}

\begin{table*}
	\caption{Numbers of GW standard sirens in cosmological analysis, triggered by THESEUS in synergy with ET, CE, 2CE, and ET2CE, respectively.}
	\label{tabsts} 
	\centering
	\setlength{\tabcolsep}{2.5mm}
	\renewcommand{\arraystretch}{2}
	\begin{tabular}{|c|*{4}{>{\centering\arraybackslash}m{1.75cm}|}}\hline
		Detection strategy&ET&CE&2CE&ET2CE\\ \hline
		Number of GW standard sirens&      400      &   538   &    600  &  640  \\    \hline
	\end{tabular}
\end{table*}

The primary goal of this work is to assess the influence of joint observations between 3G GW detectors and future GRB detectors on the cosmological measurement of sterile neutrino parameters. Such observations are crucial for alleviating the degeneracies in cosmological parameters that are commonly observed in traditional EM data. To elucidate this, we have conducted simulations to generate mock GW data, which we have subsequently integrated with the mainstream EM observations, i.e., CMB+BAO+SN data. Our analysis specifically focuses on the estimation errors and precision of sterile neutrino parameters derived from this combined dataset.

To avoid any inconsistencies in the cosmological parameters constrained by combining CMB+BAO+SN with GW mock data, we have adopted a strategic approach. This approach is designed to thoroughly investigate the capacity of GW mock data to break the parameter degeneracies present in conventional EM observations. For this purpose, we have utilized the best-fit values of the cosmological parameters derived from the CMB+BAO+SN dataset as the reference values for simulating the GW mock data corresponding to each cosmological model. This methodology enables a more accurate assessment of the role of GW data in enhancing the precision of parameter estimation and resolving the degeneracies encountered in traditional EM observations.


For simplicity, we use ``CBS" to denote the joint CMB+BAO+SN data combination. Thus, in our analysis, we use five data combinations: (1) CBS, (2) CBS+ET, (3) CBS+CE, (4) CBS+2CE, and (5) CBS+ET2CE. We will report the constraint results in the next section.

\section{Results and discussion}\label{sec3}
\begin{figure*}[!htp]
\includegraphics[scale=0.7]{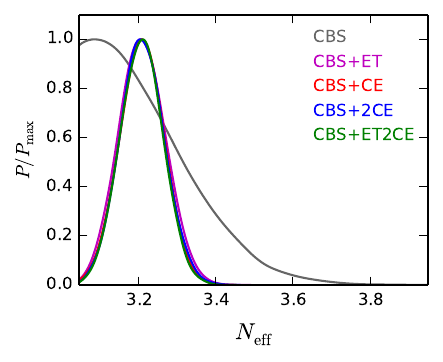}
\includegraphics[scale=0.5]{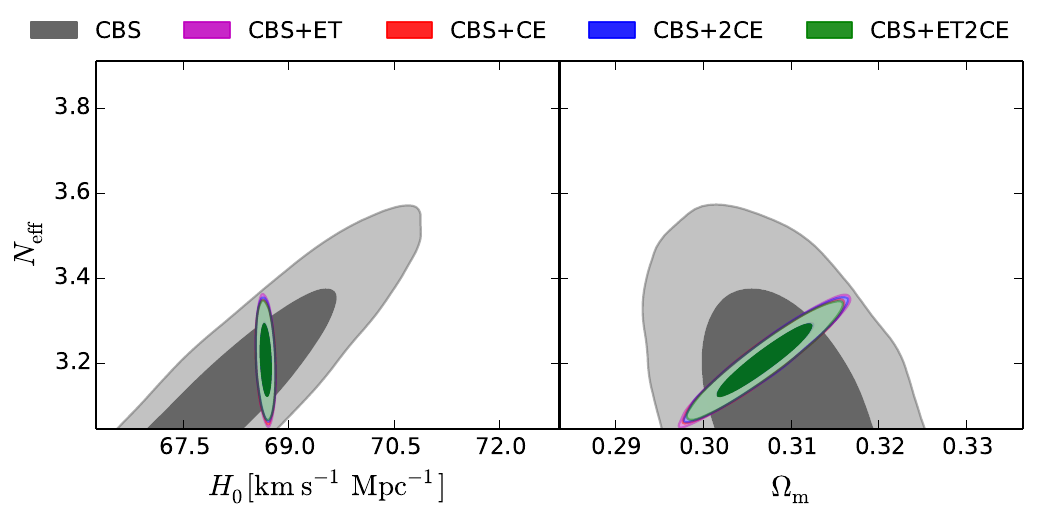}
\centering
 \caption{\label{fig3} Constraint results for the $\Lambda$CDM+$N_{\rm eff}$ model from the CBS, CBS+ET, CBS+CE, CBS+2CE, and CBS+ET2CE data combinations. One-dimensional marginalized posterior distribution for $N_{\rm eff}$ (left panel), 
and two-dimensional marginalized posterior contours (1$\sigma$ and 2$\sigma$) in the $H_0$--$N_{\rm eff}$ and $\Omega_m$--$N_{\rm eff}$ planes (right panel). }
\end{figure*}

\begin{figure*}[!htbp]
\includegraphics[width=1.0\textwidth]{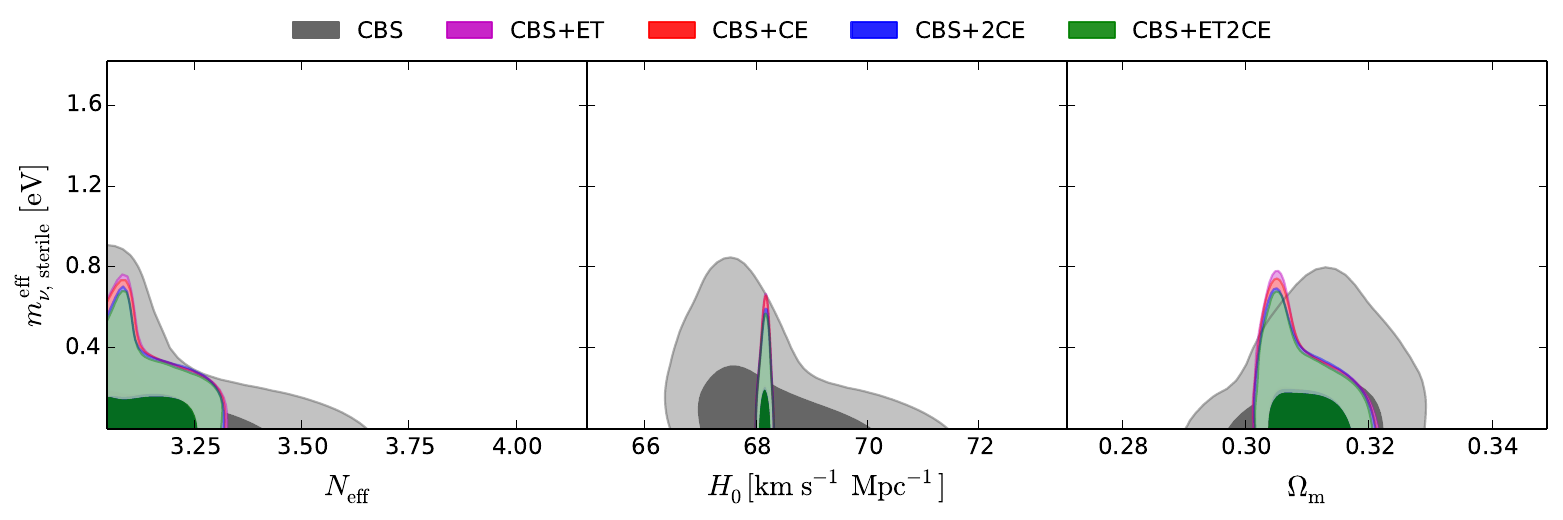}
\centering
\caption{\label{fig4}Two-dimensional marginalized posterior contours (1$\sigma$ and 2$\sigma$) in the $N_{\rm eff}$--$m_{\nu,{\rm{sterile}}}^{\rm{eff}}$, $H_0$--$m_{\nu,{\rm{sterile}}}^{\rm{eff}}$, $\Omega_m$--$m_{\nu,{\rm{sterile}}}^{\rm{eff}}$ planes for the $\Lambda$CDM+$N_{\rm eff}$+$m_{\nu,{\rm{sterile}}}^{\rm{eff}}$ model from the constraints of the CBS, CBS+ET, CBS+CE, CBS+2CE, and CBS+ET2CE data combinations.}
\end{figure*}

In this section, we report the constraint results for the $\Lambda$CDM+$N_{\rm eff}$ and $\Lambda$CDM+$N_{\rm eff}$+$m_{\nu,{\rm{sterile}}}^{\rm{eff}}$ models using the CBS, CBS+ET, CBS+CE, CBS+2CE, and CBS+ET2CE data combinations and analyze how the GW standard sirens affects the cosmological constraints on the sterile neutrino parameters.
The fitting results are shown in Figs.~\ref{fig3} and \ref{fig4} and Tables~\ref{tabless} and \ref{tabms}. In the tables, we quote $\pm 1\sigma$ errors for the parameters, but for the parameters that cannot be well constrained, e.g., the sterile neutrino parameters $N_{\rm eff}$ and $m_{\nu,{\rm{sterile}}}^{\rm{eff}}$, we quote the $2\sigma$ upper limits. For a parameter $\xi$, we use $\sigma(\xi)$ and $\varepsilon(\xi)=\sigma(\xi)/\xi$ to represent its absolute error and relative error, respectively.

In accordance with the cosmological parameter constraints obtained from the CBS data, the central values and uncertainties of the cosmological parameters from the combined CBS and GW observations are detailed in Tables \ref{tabless} and \ref{tabms}. Given that the CBS data are real, whereas the GW data are simulated, the constraints from the joint CBS and GW analysis represent a mixture of real and simulated data. Consequently, the central values should not be interpreted as from actual observational data. As such, this work emphasizes the significance of the uncertainties and precision in the derived cosmological constraints.


\begin{table*}\small
\setlength\tabcolsep{1.5pt}
\renewcommand{\arraystretch}{1.5}
\caption{\label{tabless}Fitting results of the $\Lambda$CDM+$N_{\rm eff}$ model by using the CBS, CBS+ET, CBS+CE, CBS+2CE, and CBS+ET2CE data combinations. We quote $\pm 1\sigma$ errors for the parameters, but for the parameters that cannot be well constrained, we quote the $2\sigma$ upper limits. Here, $H_0$ is in units of ${\rm km}\ {\rm s^{-1}}\  {\rm Mpc^{-1}}$.}
\centering
\begin{tabular}{ccccccccccccccccccc}

\hline
  Model && CBS && CBS+ET && CBS+CE  &&  CBS+2CE && CBS+ET2CE\\
\hline

$\Omega_bh^2$&&$0.02247\pm0.00015$&&$0.02248\pm0.00011$&&$0.02248\pm0.00011$&&$0.02248\pm0.00011$&&$0.02248\pm0.00011$\\
$\Omega_ch^2$&&$0.1220^{+0.0017}_{-0.0028}$&&$0.1217\pm0.0017$&&$0.1217\pm0.0016$&&$0.1217\pm0.0016$&&$0.1217\pm0.0016$\\
$100\theta_{MC}$&&$1.04046^{+0.00043}_{-0.00037}$&&$1.04050\pm0.00036$&&$1.04050^{+0.00034}_{-0.00035}$&&$1.04049\pm0.00035$&&$1.04050\pm0.00034$\\
$\tau$&&$0.0558^{+0.0074}_{-0.0083}$&&$0.0558^{+0.0075}_{-0.0082}$&&$0.0559^{+0.0074}_{-0.0082}$&&$0.0559^{+0.0075}_{-0.0082}$&&$0.0558^{+0.0072}_{-0.0082}$\\
$n_s$&&$0.9697^{+0.0046}_{-0.0057}$&&$0.9696\pm0.0030$&&$0.9696^{+0.0030}_{-0.0031}$&&$0.9697\pm0.0030$&&$0.9697\pm0.0029$\\
${\rm{ln}}(10^{10}A_s)$&&$3.051^{+0.016}_{-0.018}$&&$3.050\pm0.016$&&$3.050\pm0.016$&&$3.050\pm0.016$&&$3.050\pm0.016$\\
$\sigma_8$&&$0.8177^{+0.0083}_{-0.0101}$&&$0.8170^{+0.0082}_{-0.0081}$&&$0.8169\pm0.0078$&&$0.8171\pm0.0079$&&$0.8169\pm0.0077$\\
$\Omega_m$&&$0.3077^{+0.0060}_{-0.0061}$&&$0.3071\pm0.0038$&&$0.3071^{+0.0035}_{-0.0036}$&&$0.3071\pm0.0035$&&$0.3070^{+0.0034}_{-0.0035}$\\
$H_0$&&$68.670^{+0.640}_{-1.020}$&&$68.671^{+0.057}_{-0.055}$&&$68.671^{+0.056}_{-0.057}$&&$68.671^{+0.052}_{-0.054}$&&$68.672^{+0.052}_{-0.050}$\\
$N_{\rm eff}$&&$<3.464$&&$3.209\pm0.060$&&$3.209\pm0.056$&&$3.210^{+0.055}_{-0.056}$&&$3.208^{+0.054}_{-0.055}$\\
\hline
$\Delta N_{\rm eff}>0$&&$...$&&$2.717\sigma$&&$2.911\sigma$&&$2.929\sigma$&&$2.945\sigma$\\
$\sigma(\Omega_m)$&&$0.00605$&&$0.00380$&&$0.00355$&&$0.00350$&&$0.00345$\\
$\sigma(H_0)$&&$0.8300$&&$0.0560$&&$0.0565$&&$0.0530$&&$0.0510$\\
$\varepsilon(\Omega_m)$&& $1.966\%$&&$1.237\%$ && $1.156\%$&&$1.140\%$ && $1.124\%$\\
$\varepsilon(H_0)$&& $1.209\%$&&$0.082\%$ && $0.082\%$&&$0.077\%$ && $0.074\%$\\
\hline
\end{tabular}
\end{table*}

\begin{table*}\small
\setlength\tabcolsep{1.5pt}
\renewcommand{\arraystretch}{1.5}
\caption{\label{tabms}Fitting results of the $\Lambda$CDM+$N_{\rm eff}$+$m_{\nu,{\rm{sterile}}}^{\rm{eff}}$ model by using the CBS, CBS+ET, CBS+CE, CBS+2CE, and CBS+ET2CE data combinations. We quote $\pm 1\sigma$ errors for the parameters, but for the parameters that cannot be well constrained, we quote the $2\sigma$ upper limits. Here, $H_0$ is in units of ${\rm km}\ {\rm s^{-1}}\ {\rm Mpc^{-1}}$ and $m_{\nu,{\rm{sterile}}}^{\rm{eff}}$ is in units of eV.}
\centering
\begin{tabular}{ccccccccccccccccccc}

\hline
  Model && CBS && CBS+ET && CBS+CE  &&  CBS+2CE && CBS+ET2CE\\
\hline

$\Omega_bh^2$&&$0.02247^{+0.00015}_{-0.00016}$&&$0.02249\pm0.00012$&&$0.02249\pm0.00012$&&$0.02249\pm0.00012$&&$0.02249\pm0.00012$\\
$\Omega_ch^2$&&$0.1198^{+0.0036}_{-0.0031}$&&$0.1193^{+0.0033}_{-0.0018}$&&$0.1193^{+0.0032}_{-0.0017}$&&$0.1195^{+0.0031}_{-0.0017}$&&$0.1194^{+0.0030}_{-0.0017}$\\
$100\theta_{MC}$&&$1.04059^{+0.00047}_{-0.00033}$&&$1.04071^{+0.00041}_{-0.00035}$&&$1.04071^{+0.00040}_{-0.00035}$&&$1.04070^{+0.00040}_{-0.00036}$&&$1.04071^{+0.00039}_{-0.00036}$\\
$\tau$&&$0.0561^{+0.0073}_{-0.0083}$&&$0.0566^{+0.0074}_{-0.0082}$&&$0.0566^{+0.0074}_{-0.0082}$&&$0.0568^{+0.0074}_{-0.0083}$&&$0.0569^{+0.0074}_{-0.0083}$\\
$n_s$&&$0.9681^{+0.0047}_{-0.0065}$&&$0.9684\pm0.0033$&&$0.9684\pm0.0033$&&$0.9684\pm0.0033$&&$0.9684^{+0.0033}_{-0.0032}$\\
${\rm{ln}}(10^{10}A_s)$&&$3.048^{+0.016}_{-0.018}$&&$3.048^{+0.016}_{-0.017}$&&$3.048^{+0.016}_{-0.017}$&&$3.048\pm0.016$&&$3.048\pm0.016$\\
$\sigma_8$&&$0.796^{+0.024}_{-0.014}$&&$0.795^{+0.022}_{-0.012}$&&$0.795^{+0.022}_{-0.012}$&&$0.796^{+0.021}_{-0.012}$&&$0.796^{+0.022}_{-0.012}$\\
$\Omega_m$&&$0.3114\pm0.0064$&&$0.3096^{+0.0032}_{-0.0052}$&&$0.3096^{+0.0032}_{-0.0050}$&&$0.3097^{+0.0033}_{-0.0049}$&&$0.3096^{+0.0033}_{-0.0048}$\\
$H_0$&&$68.150^{+0.500}_{-1.000}$&&$68.156^{+0.055}_{-0.053}$&&$68.156^{+0.054}_{-0.053}$&&$68.156^{+0.054}_{-0.053}$&&$68.156^{+0.054}_{-0.050}$\\
$N_{\rm eff}$&&$<3.446$&&$3.148^{+0.039}_{-0.094}$&&$3.148^{+0.040}_{-0.090}$&&$3.150^{+0.043}_{-0.089}$&&$3.150^{+0.042}_{-0.089}$\\
$m_{\nu,{\rm{sterile}}}^{\rm{eff}}$&&$<0.5789$&&$<0.4842$&&$<0.4772$&&$<0.4321$&&$<0.4226$\\
\hline
$\Delta N_{\rm eff}>0$&&$...$&&$1.085\sigma$&&$1.133\sigma$&&$1.169\sigma$&&$1.169\sigma$\\
$\sigma(\Omega_m)$&&$0.00640$&&$0.00420$&&$0.00410$&&$0.00410$&&$0.00405$\\
$\sigma(H_0)$&&$0.7500$&&$0.0540$&&$0.0535$&&$0.0535$&&$0.0520$\\
$\varepsilon(\Omega_m)$&& $2.055\%$&&$1.357\%$ && $1.324\%$&&$1.324\%$ && $1.308\%$\\
$\varepsilon(H_0)$&& $1.101\%$&&$0.079\%$ && $0.078\%$&&$0.078\%$ && $0.076\%$\\
\hline
\end{tabular}
\end{table*}

\subsection{The case of massless sterile neutrino}
The massless sterile neutrinos serve as the dark radiation, and thus in this case $N_{\rm eff}$ is treated as a free parameter, the total relativistic energy density of radiation is given by
$$\rho_{\rm r}=[1+N_{\rm eff}\frac{7}{8}(\frac{4}{11})^{\frac{4}{3}}]\rho_\gamma,$$
where $\rho_\gamma$ is the photon energy density. In the $\Lambda$CDM model, $N_{\rm eff}=3.046$, and so $\Delta N_{\rm eff}=N_{\rm eff}-3.046>0$ indicates the presence of extra relativistic particle species in the early universe, and in this paper we take the fit results of $\Delta N_{\rm eff}>0$ as evidence of the existence of massless sterile neutrinos.

In Table~\ref{tabless}, we find that the CBS data provides only an upper limit, $N_{\rm eff}<3.464$, indicating that the existence of massless sterile neutrinos is not favored by the CBS data. 
However, when GW standard sirens are included in the data combination, the results change significantly.
After considering the GW data, we obtain $N_{\rm eff}=3.209\pm0.060$ for CBS+ET, $N_{\rm eff}=3.209\pm0.056$ for CBS+CE, $N_{\rm eff}=3.210^{+0.055}_{-0.056}$ for CBS+2CE, and $N_{\rm eff}=3.208^{+0.054}_{-0.055}$ for CBS+ET2CE, which indicates a detection of $\Delta N_{\rm eff}>0$ at the $2.717\sigma$, $2.911\sigma$, $2.929\sigma$, and $2.945\sigma$ significance levels, respectively.
Obviously, the GW data can indeed effectively improve the constraints on the massless sterile neutrino parameter $N_{\rm eff}$. 

Note that here CBS refers to current real observational data, while GW denotes simulated data. When we use the actual CBS data to conduct constraints, we can only obtain an upper limit for $N_{\rm eff}$ and not a definitive detection result. However, when simulated GW data are included, the errors in the parameter constraints are significantly reduced. With the central value of $N_{\rm eff}$ remaining essentially unchanged and $\Delta N_{\rm eff}$ substantially decreased, we can achieve a result of $\Delta N_{\rm eff}> 0$. Our results indicate that future GW observations can greatly assist in improving the cosmological detection of massless sterile neutrinos, with a significance level reaching up to $3\sigma$. Of course, we must remember that our findings are based on simulated outcomes, and the central value of $N_{\rm eff}$ is set to be consistent with the result from CBS.



In the right panel of Fig.~\ref{fig3}, we show the two-dimensional posterior distribution contours (1$\sigma$ and 2$\sigma$) in the $N_{\rm eff}$--$H_0$ and $N_{\rm eff}$--$\Omega_m$ planes for the $\Lambda$CDM+$N_{\rm eff}$ model using CBS, CBS+ET, CBS+CE, CBS+2CE, and CBS+ET2CE data combinations. 
We can see that $N_{\rm eff}$ is in positive correlation with $H_0$ by using the CBS data combination. However, after adding the GW data, this correlation becomes negligible, indicating that the degeneracy between $N_{\rm eff}$ and $H_0$ is effectively broken by the GW observations.
In addition, we can also clearly see that when considering the GW data, the parameter space is greatly shrunck in each planes and the constraints on cosmological parameters of $H_0$ and $\Omega_m$ become much tighter.

In Table~\ref{tabless}, we also show absolute and relative errors of $H_0$ and $\Omega_m$ from the CBS, CBS+ET, CBS+CE, CBS+2CE, and CBS+ET2CE data combinations. Compared to the CBS data, we find that the accuracy of the $H_0$ constraint improves by $93.253\%$ for CBS+ET, $93.193\%$ for CBS+CE, $93.614\%$ for CBS+2CE, and $93.855\%$ for CBS+ET2CE, respectively. Similarly, the accuracy of the $\Omega_m$ constraint improves by $37.190\%$ for CBS+ET, $41.322\%$ for CBS+CE, $42.149\%$ for CBS+2CE, and $42.975\%$ for CBS+ET2CE, respectively.
Obviously, the GW data can indeed effectively improve the constraints on the parameters $H_0$ and $\Omega_m$.

\subsection{The case of massive sterile neutrino}
In this subsection, we investigated how GW standard sirens on constraint the massive sterile neutrino parameters. Hence, the requirement of $N_{\rm eff}>3.046$ still holds.

From Table~\ref{tabms}, we obtain $N_{\rm eff}<3.446$ by using CBS data. After adding the GW data, the constraint results become $N_{\rm eff}=3.148^{+0.039}_{-0.094}$ for CBS+ET, $N_{\rm eff}=3.148^{+0.039}_{-0.090}$ for CBS+CE, $N_{\rm eff}=3.150^{+0.043}_{-0.089}$ for CBS+2CE, and $N_{\rm eff}=3.150^{+0.042}_{-0.089}$ for CBS+ET2CE, respectively.
We find that $N_{\rm eff}$ cannot be well constrained using the CBS data, but the addition of GW data can significantly improve the constraint on $N_{\rm eff}$, favoring $\Delta N_{\rm eff}>0$ at $1.085\sigma$ (CBS+ET), $1.133\sigma$ (CBS+CE), $1.169\sigma$ (CBS+2CE), and $1.169\sigma$ (CBS+ET2CE) statistical significance, respectively.
For the mass of sterile neutrino, the CBS data give $m_{\nu,{\rm{sterile}}}^{\rm{eff}}<0.5789$ eV and further including the GW data leads to results of $m_{\nu,{\rm{sterile}}}^{\rm{eff}}<0.4842$ eV (CBS+ET), $m_{\nu,{\rm{sterile}}}^{\rm{eff}}<0.4772$ eV (CBS+CE), $m_{\nu,{\rm{sterile}}}^{\rm{eff}}<0.4321$ eV (CBS+2CE), and $m_{\nu,{\rm{sterile}}}^{\rm{eff}}<0.4226$ eV (CBS+ET2CE), respectively.
Evidently, adding GW data tightens the constraint on $m_{\nu,{\rm{sterile}}}^{\rm{eff}}$ significantly, which is in accordance with the conclusions in previous studies active neutrinos mass by using the GW data~\cite{Wang:2018lun,Jin:2022tdf,Feng:2024lzh}.
Therefore, the GW data also play an important role in constraining the mass of sterile neutrinos.

In Fig.~\ref{fig4}, we show two-dimensional marginalized posterior contours (1$\sigma$ and 2$\sigma$) in the $N_{\rm eff}$--$m_{\nu,{\rm{sterile}}}^{\rm{eff}}$, $H_0$--$m_{\nu,{\rm{sterile}}}^{\rm{eff}}$, $\Omega_m$--$m_{\nu,{\rm{sterile}}}^{\rm{eff}}$ planes for the $\Lambda$CDM+$N_{\rm eff}$+$m_{\nu,{\rm{sterile}}}^{\rm{eff}}$ model.
We can clearly see that when further considering the GW data, the parameter space is also greatly shrunk and the constraints on $H_0$ and $\Omega_m$ also become much tighter.
From Table~\ref{tabms}, we find that the constraints on $H_0$ and $\Omega_m$ could be improved by $92.800\%$ and $34.375\%$, respectively, when adding the ET data to the CBS data, $92.867\%$ and $35.938\%$ for the case of CE, $92.867\%$ and $35.938\%$ for the case of 2CE, and $93.067\%$ and $36.719\%$ for the case of ET2CE, respectively.
These result in accordance with the conclusions for both the case of massless sterile neutrinos in this study and active neutrinos mass in previous studies~\cite{Wang:2018lun,Jin:2022tdf,Feng:2024lzh}.
Therefore, the inclusion of GW data can significantly improve constraints on most cosmological parameters, particularly $H_0$ and $\Omega_m$.

\section{Conclusion}\label{sec4}
This work aims to forecast the search for sterile neutrinos using joint GW-GRB observations. We consider two cases of massless and massive sterile neutrinos, corresponding to the $\Lambda$CDM+$N_{\rm eff}$ and $\Lambda$CDM+$N_{\rm eff}$+$m_{\nu,{\rm{sterile}}}^{\rm{eff}}$ models, respectively.
We consider four GW detection observation strategies, i.e., ET, CE, the 2CE network, and the ET2CE network, to perform cosmological analysis.
To evaluate the impact of GW data on the constraints of sterile neutrino parameters, we also considered existing CMB+BAO+SN data for comparison and combination.

For the $\Lambda$CDM+$N_{\rm eff}$ model, in the case of using CMB+BAO+SN, only upper limits on $N_{\rm eff}$ can be obtained. Further adding the GW data tightens the $N_{\rm eff}$ significantly, and in this case the preference of $\Delta N_{\rm eff}>0$ at about $3\sigma$ level.
Therefore, GW standard siren observations can greatly assist in the detection of massless sterile neutrinos.


For the $\Lambda$CDM+$N_{\rm eff}$+$m_{\nu,{\rm{sterile}}}^{\rm{eff}}$ model, only upper limits on $N_{\rm eff}$ and $m_{\nu,{\rm{sterile}}}^{\rm{eff}}$ can be derived by using the CMB+BAO+SN data. Further including GW data significantly improves the constraints, and we find that the GW data give a rather tight upper limit on $m_{\nu,{\rm{sterile}}}^{\rm{eff}}$ and favor $\Delta N_{\rm eff}>0$ at about $1.1\sigma$ level. This results also seems to favor a massless sterile neutrinos.

Furthermore, we find that the GW data can significantly enhances the accuracy of constraints on the derived parameters $H_0$ and $\Omega_m$. The accuracy of $H_0$ improves by approximately $93\%$ and the accuracy of $\Omega_m$ increases by about $37\%$ to $42\%$, when the GW data are included in the cosmological fit.

\begin{acknowledgments}
This work was supported by the National Natural Science Foundation of China (Grant Nos. 12305069, 11947022, 12473001, 11975072, 11875102, and 11835009), the National SKA Program of China (Grants Nos. 2022SKA0110200 and 2022SKA0110203), the Program of the Education Department of Liaoning Province (Grant No. JYTMS20231695), and the National 111 Project (Grant No. B16009).

\end{acknowledgments}
\bibliographystyle{unsrt}
\bibliography{sterile}

\begin{thebibliography}{100}

\bibitem{LIGOScientific:2017vwq}
B.~P. Abbott et~al.
\newblock {GW170817: Observation of Gravitational Waves from a Binary Neutron
  Star Inspiral}.
\newblock {\em Phys. Rev. Lett.}, 119(16):161101, 2017.

\bibitem{LIGOScientific:2017ync}
B.~P. Abbott et~al.
\newblock {Multi-messenger Observations of a Binary Neutron Star Merger}.
\newblock {\em Astrophys. J. Lett.}, 848(2):L12, 2017.

\bibitem{TOROS:2017pqe}
M.~C. D\'\i{}az et~al.
\newblock {Observations of the first electromagnetic counterpart to a
  gravitational wave source by the TOROS collaboration}.
\newblock {\em Astrophys. J. Lett.}, 848(2):L29, 2017.

\bibitem{ET-web}
{\rm ET}.
\newblock \url{https://www.et-gw.eu/}.

\bibitem{Punturo:2010zz}
M.~Punturo et~al.
\newblock {The Einstein Telescope: A third-generation gravitational wave
  observatory}.
\newblock {\em Class. Quant. Grav.}, 27:194002, 2010.

\bibitem{CE-web}
{\rm CE}.
\newblock \url{https://cosmicexplorer.org/}.

\bibitem{Evans:2016mbw}
Benjamin~P Abbott et~al.
\newblock {Exploring the Sensitivity of Next Generation Gravitational Wave
  Detectors}.
\newblock {\em Class. Quant. Grav.}, 34(4):044001, 2017.

\bibitem{Cai:2016sby}
Rong-Gen Cai and Tao Yang.
\newblock {Estimating cosmological parameters by the simulated data of
  gravitational waves from the Einstein Telescope}.
\newblock {\em Phys. Rev. D}, 95(4):044024, 2017.

\bibitem{Cai:2017plb}
Rong-Gen Cai and Tao Yang.
\newblock {Standard sirens and dark sector with Gaussian process}.
\newblock {\em EPJ Web Conf.}, 168:01008, 2018.

\bibitem{Liu:2017xef}
Tan Liu, Xing Zhang, and Wen Zhao.
\newblock {Constraining $f(R)$ gravity in solar system, cosmology and binary
  pulsar systems}.
\newblock {\em Phys. Lett. B}, 777:286--293, 2018.

\bibitem{Cai:2017aea}
Rong-Gen Cai, Tong-Bo Liu, Xue-Wen Liu, Shao-Jiang Wang, and Tao Yang.
\newblock {Probing cosmic anisotropy with gravitational waves as standard
  sirens}.
\newblock {\em Phys. Rev. D}, 97(10):103005, 2018.

\bibitem{Berti:2018cxi}
Emanuele Berti, Kent Yagi, and Nicol\'as Yunes.
\newblock {Extreme Gravity Tests with Gravitational Waves from Compact Binary
  Coalescences: (I) Inspiral-Merger}.
\newblock {\em Gen. Rel. Grav.}, 50(4):46, 2018.

\bibitem{Cai:2018rzd}
Yi-Fu Cai, Chunlong Li, Emmanuel~N. Saridakis, and Lingqin Xue.
\newblock {$f(T)$ gravity after GW170817 and GRB170817A}.
\newblock {\em Phys. Rev. D}, 97(10):103513, 2018.

\bibitem{Wang:2018lun}
Ling-Feng Wang, Xuan-Neng Zhang, Jing-Fei Zhang, and Xin Zhang.
\newblock {Impacts of gravitational-wave standard siren observation of the
  Einstein Telescope on weighing neutrinos in cosmology}.
\newblock {\em Phys. Lett. B}, 782:87--93, 2018.

\bibitem{Zhao:2018gwk}
Wen Zhao, Bill~S. Wright, and Baojiu Li.
\newblock {Constraining the time variation of Newton's constant $G$ with
  gravitational-wave standard sirens and supernovae}.
\newblock {\em JCAP}, 10:052, 2018.

\bibitem{Zhang:2018byx}
Xuan-Neng Zhang, Ling-Feng Wang, Jing-Fei Zhang, and Xin Zhang.
\newblock {Improving cosmological parameter estimation with the future
  gravitational-wave standard siren observation from the Einstein Telescope}.
\newblock {\em Phys. Rev. D}, 99(6):063510, 2019.

\bibitem{Du:2018tia}
Minghui Du, Weiqiang Yang, Lixin Xu, Supriya Pan, and David~F. Mota.
\newblock {Future constraints on dynamical dark-energy using gravitational-wave
  standard sirens}.
\newblock {\em Phys. Rev. D}, 100(4):043535, 2019.

\bibitem{He:2019dhl}
Jian-hua He.
\newblock {Accurate method to determine the systematics due to the peculiar
  velocities of galaxies in measuring the Hubble constant from
  gravitational-wave standard sirens}.
\newblock {\em Phys. Rev. D}, 100(2):023527, 2019.

\bibitem{Yang:2019bpr}
Weiqiang Yang, Supriya Pan, Eleonora Di~Valentino, Bin Wang, and Anzhong Wang.
\newblock {Forecasting interacting vacuum-energy models using gravitational
  waves}.
\newblock {\em JCAP}, 05:050, 2020.

\bibitem{Yang:2019vni}
Weiqiang Yang, Sunny Vagnozzi, Eleonora Di~Valentino, Rafael~C. Nunes, Supriya
  Pan, and David~F. Mota.
\newblock {Listening to the sound of dark sector interactions with
  gravitational wave standard sirens}.
\newblock {\em JCAP}, 07:037, 2019.

\bibitem{Zhang:2019ylr}
Xin Zhang.
\newblock {Gravitational wave standard sirens and cosmological parameter
  measurement}.
\newblock {\em Sci. China Phys. Mech. Astron.}, 62(11):110431, 2019.

\bibitem{Zhang:2019ple}
Jing-Fei Zhang, Hong-Yan Dong, Jing-Zhao Qi, and Xin Zhang.
\newblock {Prospect for constraining holographic dark energy with gravitational
  wave standard sirens from the Einstein Telescope}.
\newblock {\em Eur. Phys. J. C}, 80(3):217, 2020.

\bibitem{Bachega:2019fki}
Riis R.~A. Bachega, Andr\'e~A. Costa, E.~Abdalla, and K.~S.~F. Fornazier.
\newblock {Forecasting the Interaction in Dark Matter-Dark Energy Models with
  Standard Sirens From the Einstein Telescope}.
\newblock {\em JCAP}, 05:021, 2020.

\bibitem{Wang:2019tto}
Ling-Feng Wang, Ze-Wei Zhao, Jing-Fei Zhang, and Xin Zhang.
\newblock {A preliminary forecast for cosmological parameter estimation with
  gravitational-wave standard sirens from TianQin}.
\newblock {\em JCAP}, 11:012, 2020.

\bibitem{Zhang:2019loq}
Jing-Fei Zhang, Ming Zhang, Shang-Jie Jin, Jing-Zhao Qi, and Xin Zhang.
\newblock {Cosmological parameter estimation with future gravitational wave
  standard siren observation from the Einstein Telescope}.
\newblock {\em JCAP}, 09:068, 2019.

\bibitem{Li:2019ajo}
Hai-Li Li, Dong-Ze He, Jing-Fei Zhang, and Xin Zhang.
\newblock {Quantifying the impacts of future gravitational-wave data on
  constraining interacting dark energy}.
\newblock {\em JCAP}, 06:038, 2020.

\bibitem{Zhao:2019gyk}
Ze-Wei Zhao, Ling-Feng Wang, Jing-Fei Zhang, and Xin Zhang.
\newblock {Prospects for improving cosmological parameter estimation with
  gravitational-wave standard sirens from Taiji}.
\newblock {\em Sci. Bull.}, 65(16):1340--1348, 2020.

\bibitem{Jin:2020hmc}
Shang-Jie Jin, Dong-Ze He, Yidong Xu, Jing-Fei Zhang, and Xin Zhang.
\newblock {Forecast for cosmological parameter estimation with
  gravitational-wave standard siren observation from the Cosmic Explorer}.
\newblock {\em JCAP}, 03:051, 2020.

\bibitem{Wang:2021srv}
Ling-Feng Wang, Shang-Jie Jin, Jing-Fei Zhang, and Xin Zhang.
\newblock {Forecast for cosmological parameter estimation with
  gravitational-wave standard sirens from the LISA-Taiji network}.
\newblock {\em Sci. China Phys. Mech. Astron.}, 65(1):210411, 2022.

\bibitem{Qi:2021iic}
Jing-Zhao Qi, Shang-Jie Jin, Xi-Long Fan, Jing-Fei Zhang, and Xin Zhang.
\newblock {Using a multi-messenger and multi-wavelength observational strategy
  to probe the nature of dark energy through direct measurements of cosmic
  expansion history}.
\newblock {\em JCAP}, 12(12):042, 2021.

\bibitem{Jin:2021pcv}
Shang-Jie Jin, Ling-Feng Wang, Peng-Ju Wu, Jing-Fei Zhang, and Xin Zhang.
\newblock {How can gravitational-wave standard sirens and 21-cm intensity
  mapping jointly provide a precise late-universe cosmological probe?}
\newblock {\em Phys. Rev. D}, 104(10):103507, 2021.

\bibitem{Zhu:2021bpp}
Liang-Gui Zhu, Ling-Hua Xie, Yi-Ming Hu, Shuai Liu, En-Kun Li, Nicola~R.
  Napolitano, Bai-Tian Tang, Jian-dong Zhang, and Jianwei Mei.
\newblock {Constraining the Hubble constant to a precision of about 1\% using
  multi-band dark standard siren detections}.
\newblock {\em Sci. China Phys. Mech. Astron.}, 65(5):259811, 2022.

\bibitem{deSouza:2021xtg}
Josiel Mendon\c{c}a~Soares de~Souza, Riccardo Sturani, and Jailson Alcaniz.
\newblock {Cosmography with standard sirens from the Einstein Telescope}.
\newblock {\em JCAP}, 03(03):025, 2022.

\bibitem{Wang:2022oou}
Ling-Feng Wang, Yue Shao, Jing-Fei Zhang, and Xin Zhang.
\newblock {Ultra-low-frequency gravitational waves from individual supermassive
  black hole binaries as standard sirens}.
\newblock 1 2022.

\bibitem{Wu:2022dgy}
Peng-Ju Wu, Yue Shao, Shang-Jie Jin, and Xin Zhang.
\newblock {A path to precision cosmology: synergy between four promising
  late-universe cosmological probes}.
\newblock {\em JCAP}, 06:052, 2023.

\bibitem{Jin:2022qnj}
Shang-Jie Jin, Tian-Nuo Li, Jing-Fei Zhang, and Xin Zhang.
\newblock {Prospects for measuring the Hubble constant and dark energy using
  gravitational-wave dark sirens with neutron star tidal deformation}.
\newblock {\em JCAP}, 08:070, 2023.

\bibitem{Jin:2022tdf}
Shang-Jie Jin, Rui-Qi Zhu, Ling-Feng Wang, Hai-Li Li, Jing-Fei Zhang, and Xin
  Zhang.
\newblock {Impacts of gravitational-wave standard siren observations from
  Einstein Telescope and Cosmic Explorer on weighing neutrinos in interacting
  dark energy models}.
\newblock {\em Commun. Theor. Phys.}, 74(10):105404, 2022.

\bibitem{Hou:2022rvk}
Wan-Ting Hou, Jing-Zhao Qi, Tao Han, Jing-Fei Zhang, Shuo Cao, and Xin Zhang.
\newblock {Prospects for constraining interacting dark energy models from
  gravitational wave and gamma ray burst joint observation}.
\newblock {\em JCAP}, 05:017, 2023.

\bibitem{Song:2022siz}
Ji-Yu Song, Ling-Feng Wang, Yichao Li, Ze-Wei Zhao, Jing-Fei Zhang, Wen Zhao,
  and Xin Zhang.
\newblock {Synergy between CSST galaxy survey and gravitational-wave
  observation: Inferring the Hubble constant from dark standard sirens}.
\newblock {\em Sci. China Phys. Mech. Astron.}, 67(3):230411, 2024.

\bibitem{Jin:2023zhi}
Shang-Jie Jin, Shuang-Shuang Xing, Yue Shao, Jing-Fei Zhang, and Xin Zhang.
\newblock {Joint constraints on cosmological parameters using future multi-band
  gravitational wave standard siren observations*}.
\newblock {\em Chin. Phys. C}, 47(6):065104, 2023.

\bibitem{Jin:2023sfc}
Shang-Jie Jin, Ye-Zhu Zhang, Ji-Yu Song, Jing-Fei Zhang, and Xin Zhang.
\newblock {Taiji-TianQin-LISA network: Precisely measuring the Hubble constant
  using both bright and dark sirens}.
\newblock {\em Sci. China Phys. Mech. Astron.}, 67(2):220412, 2024.

\bibitem{Jin:2023tou}
Shang-Jie Jin, Rui-Qi Zhu, Ji-Yu Song, Tao Han, Jing-Fei Zhang, and Xin Zhang.
\newblock {Standard siren cosmology in the era of the 2.5-generation
  ground-based gravitational wave detectors: bright and dark sirens of LIGO
  Voyager and NEMO}.
\newblock 9 2023.

\bibitem{Han:2023exn}
Tao Han, Shang-Jie Jin, Jing-Fei Zhang, and Xin Zhang.
\newblock {A comprehensive forecast for cosmological parameter estimation using
  joint observations of gravitational waves and short $\gamma $-ray bursts}.
\newblock {\em Eur. Phys. J. C}, 84(7):663, 2024.

\bibitem{Li:2023gtu}
Tian-Nuo Li, Shang-Jie Jin, Hai-Li Li, Jing-Fei Zhang, and Xin Zhang.
\newblock {Prospects for Probing the Interaction between Dark Energy and Dark
  Matter Using Gravitational-wave Dark Sirens with Neutron Star Tidal
  Deformation}.
\newblock {\em Astrophys. J.}, 963(1):52, 2024.

\bibitem{Dong:2024bvw}
Yue-Yan Dong, Ji-Yu Song, Shang-Jie Jin, Jing-Fei Zhang, and Xin Zhang.
\newblock {Enhancing dark siren cosmology through multi-band gravitational wave
  synergetic observations}.
\newblock 4 2024.

\bibitem{Feng:2024lzh}
Lu~Feng, Tao Han, Jing-Fei Zhang, and Xin Zhang.
\newblock {Prospects for weighing neutrinos in interacting dark energy models
  using joint observations of gravitational waves and $\gamma$-ray bursts}.
\newblock {\em Chin. Phys. C}, 48:095104, 2024.

\bibitem{Bian:2021ini}
Ligong Bian et~al.
\newblock {The Gravitational-wave physics II: Progress}.
\newblock {\em Sci. China Phys. Mech. Astron.}, 64(12):120401, 2021.

\bibitem{LSND:2001aii}
A.~Aguilar et~al.
\newblock {Evidence for neutrino oscillations from the observation of
  $\bar{\nu}_e$ appearance in a $\bar{\nu}_\mu$ beam}.
\newblock {\em Phys. Rev. D}, 64:112007, 2001.

\bibitem{Giunti:2010zu}
Carlo Giunti and Marco Laveder.
\newblock {Statistical Significance of the Gallium Anomaly}.
\newblock {\em Phys. Rev. C}, 83:065504, 2011.

\bibitem{Akbar:2011qw}
M.~Akbar, H.~Quevedo, K.~Saifullah, A.~Sanchez, and S.~Taj.
\newblock {Thermodynamic Geometry Of Charged Rotating BTZ Black Holes}.
\newblock {\em Phys. Rev. D}, 83:084031, 2011.

\bibitem{Conrad:2012qt}
J.~M. Conrad, C.~M. Ignarra, G.~Karagiorgi, M.~H. Shaevitz, and J.~Spitz.
\newblock {Sterile Neutrino Fits to Short Baseline Neutrino Oscillation
  Measurements}.
\newblock {\em Adv. High Energy Phys.}, 2013:163897, 2013.

\bibitem{MiniBooNE:2012maf}
A.~A. Aguilar-Arevalo et~al.
\newblock {A Combined $\nu_\mu \rightarrow \nu_e$ and $\bar \nu_\mu \rightarrow
  \bar \nu_e$ Oscillation Analysis of the MiniBooNE Excesses}.
\newblock 7 2012.

\bibitem{Giunti:2012tn}
C.~Giunti, M.~Laveder, Y.~F. Li, Q.~Y. Liu, and H.~W. Long.
\newblock {Update of Short-Baseline Electron Neutrino and Antineutrino
  Disappearance}.
\newblock {\em Phys. Rev. D}, 86:113014, 2012.

\bibitem{Giunti:2012bc}
C.~Giunti, M.~Laveder, Y.~F. Li, and H.~W. Long.
\newblock {Short-baseline electron neutrino oscillation length after troitsk}.
\newblock {\em Phys. Rev. D}, 87(1):013004, 2013.

\bibitem{Kopp:2013vaa}
Joachim Kopp, Pedro A.~N. Machado, Michele Maltoni, and Thomas Schwetz.
\newblock {Sterile Neutrino Oscillations: The Global Picture}.
\newblock {\em JHEP}, 05:050, 2013.

\bibitem{Giunti:2013aea}
C.~Giunti, M.~Laveder, Y.~F. Li, and H.~W. Long.
\newblock {Pragmatic View of Short-Baseline Neutrino Oscillations}.
\newblock {\em Phys. Rev. D}, 88:073008, 2013.

\bibitem{Gariazzo:2013gua}
S.~Gariazzo, C.~Giunti, and M.~Laveder.
\newblock {Light Sterile Neutrinos in Cosmology and Short-Baseline Oscillation
  Experiments}.
\newblock {\em JHEP}, 11:211, 2013.

\bibitem{Abazajian:2012ys}
K.~N. Abazajian et~al.
\newblock {Light Sterile Neutrinos: A White Paper}.
\newblock 4 2012.

\bibitem{Hannestad:2012ky}
Steen Hannestad, Irene Tamborra, and Thomas Tram.
\newblock {Thermalisation of light sterile neutrinos in the early universe}.
\newblock {\em JCAP}, 07:025, 2012.

\bibitem{Conrad:2013mka}
Janet~M. Conrad, William~C. Louis, and Michael~H. Shaevitz.
\newblock {The LSND and MiniBooNE Oscillation Searches at High $\Delta m^2$}.
\newblock {\em Ann. Rev. Nucl. Part. Sci.}, 63:45--67, 2013.

\bibitem{Hu:1997mj}
Wayne Hu, Daniel~J. Eisenstein, and Max Tegmark.
\newblock {Weighing neutrinos with galaxy surveys}.
\newblock {\em Phys. Rev. Lett.}, 80:5255--5258, 1998.

\bibitem{Reid:2009nq}
Beth~A. Reid, Licia Verde, Raul Jimenez, and Olga Mena.
\newblock {Robust Neutrino Constraints by Combining Low Redshift Observations
  with the CMB}.
\newblock {\em JCAP}, 01:003, 2010.

\bibitem{Li:2012vn}
Hong Li and Xin Zhang.
\newblock {Constraining dynamical dark energy with a divergence-free
  parametrization in the presence of spatial curvature and massive neutrinos}.
\newblock {\em Phys. Lett. B}, 713:160--164, 2012.

\bibitem{Wang:2012vh}
Xin Wang, Xiao-Lei Meng, Tong-Jie Zhang, HuanYuan Shan, Yan Gong, Charling Tao,
  Xuelei Chen, and Y.~F. Huang.
\newblock {Observational constraints on cosmic neutrinos and dark energy
  revisited}.
\newblock {\em JCAP}, 11:018, 2012.

\bibitem{Hamann:2012fe}
Jan Hamann, Steen Hannestad, and Yvonne Y.~Y. Wong.
\newblock {Measuring neutrino masses with a future galaxy survey}.
\newblock {\em JCAP}, 11:052, 2012.

\bibitem{Li:2012spm}
Yun-He Li, Shuang Wang, Xiao-Dong Li, and Xin Zhang.
\newblock {Holographic dark energy in a Universe with spatial curvature and
  massive neutrinos: a full Markov Chain Monte Carlo exploration}.
\newblock {\em JCAP}, 02:033, 2013.

\bibitem{Riemer-Sorensen:2013jsa}
Signe Riemer-S\o{}rensen, David Parkinson, and Tamara~M. Davis.
\newblock {Combining Planck data with large scale structure information gives a
  strong neutrino mass constraint}.
\newblock {\em Phys. Rev. D}, 89:103505, 2014.

\bibitem{Giusarma:2013pmn}
Elena Giusarma, Roland de~Putter, Shirley Ho, and Olga Mena.
\newblock {Constraints on neutrino masses from Planck and Galaxy Clustering
  data}.
\newblock {\em Phys. Rev. D}, 88(6):063515, 2013.

\bibitem{Cahn:2013taa}
R.~N. Cahn, D.~A. Dwyer, S.~J. Freedman, W.~C. Haxton, R.~W. Kadel, Yu.~G.
  Kolomensky, K.~B. Luk, P.~McDonald, G.~D. Orebi~Gann, and A.~W.~P. Poon.
\newblock {White Paper: Measuring the Neutrino Mass Hierarchy}.
\newblock In {\em {Snowmass 2013}: {Snowmass on the Mississippi}}, 7 2013.

\bibitem{Lesgourgues:2014zoa}
Julien Lesgourgues and Sergio Pastor.
\newblock {Neutrino cosmology and Planck}.
\newblock {\em New J. Phys.}, 16:065002, 2014.

\bibitem{Zhang:2014nta}
Jing-Fei Zhang, Yun-He Li, and Xin Zhang.
\newblock {Cosmological constraints on neutrinos after BICEP2}.
\newblock {\em Eur. Phys. J. C}, 74:2954, 2014.

\bibitem{Zhou:2014fva}
Xiao-Ying Zhou and Jian-Hua He.
\newblock {Weighing neutrinos in $f(R)$ gravity in light of BICEP2}.
\newblock {\em Commun. Theor. Phys.}, 62:102--108, 2014.

\bibitem{Costanzi:2014tna}
Matteo Costanzi, Barbara Sartoris, Matteo Viel, and Stefano Borgani.
\newblock {Neutrino constraints: what large-scale structure and CMB data are
  telling us?}
\newblock {\em JCAP}, 10:081, 2014.

\bibitem{Palanque-Delabrouille:2014jca}
Nathalie Palanque-Delabrouille et~al.
\newblock {Constraint on neutrino masses from SDSS-III/BOSS Ly$\alpha$ forest
  and other cosmological probes}.
\newblock {\em JCAP}, 02:045, 2015.

\bibitem{Zhang:2015rha}
Jing-Fei Zhang, Ming-Ming Zhao, Yun-He Li, and Xin Zhang.
\newblock {Neutrinos in the holographic dark energy model: constraints from
  latest measurements of expansion history and growth of structure}.
\newblock {\em JCAP}, 04:038, 2015.

\bibitem{Qian:2015waa}
X.~Qian and P.~Vogel.
\newblock {Neutrino Mass Hierarchy}.
\newblock {\em Prog. Part. Nucl. Phys.}, 83:1--30, 2015.

\bibitem{Patterson:2015xja}
R.~B. Patterson.
\newblock {Prospects for Measurement of the Neutrino Mass Hierarchy}.
\newblock {\em Ann. Rev. Nucl. Part. Sci.}, 65:177--192, 2015.

\bibitem{Allison:2015qca}
R.~Allison, P.~Caucal, E.~Calabrese, J.~Dunkley, and T.~Louis.
\newblock {Towards a cosmological neutrino mass detection}.
\newblock {\em Phys. Rev. D}, 92(12):123535, 2015.

\bibitem{Geng:2015haa}
Chao-Qiang Geng, Chung-Chi Lee, R.~Myrzakulov, M.~Sami, and Emmanuel~N.
  Saridakis.
\newblock {Observational constraints on varying neutrino-mass cosmology}.
\newblock {\em JCAP}, 01:049, 2016.

\bibitem{Chen:2015oga}
Yun Chen and Lixin Xu.
\newblock {Galaxy clustering, CMB and supernova data constraints on
  \ensuremath{\phi} CDM model with massive neutrinos}.
\newblock {\em Phys. Lett. B}, 752:66--75, 2016.

\bibitem{Zhang:2015uhk}
Xin Zhang.
\newblock {Impacts of dark energy on weighing neutrinos after Planck 2015}.
\newblock {\em Phys. Rev. D}, 93(8):083011, 2016.

\bibitem{Huang:2015wrx}
Qing-Guo Huang, Ke~Wang, and Sai Wang.
\newblock {Constraints on the neutrino mass and mass hierarchy from
  cosmological observations}.
\newblock {\em Eur. Phys. J. C}, 76(9):489, 2016.

\bibitem{Chen:2016eyp}
Yun Chen, Bharat Ratra, Marek Biesiada, Song Li, and Zong-Hong Zhu.
\newblock {Constraints on non-flat cosmologies with massive neutrinos after
  Planck 2015}.
\newblock {\em Astrophys. J.}, 829(2):61, 2016.

\bibitem{Moresco:2016nqq}
Michele Moresco, Raul Jimenez, Licia Verde, Andrea Cimatti, Lucia Pozzetti,
  Claudia Maraston, and Daniel Thomas.
\newblock {Constraining the time evolution of dark energy, curvature and
  neutrino properties with cosmic chronometers}.
\newblock {\em JCAP}, 12:039, 2016.

\bibitem{Lu:2016hsd}
Jianbo Lu, Molin Liu, Yabo Wu, Yan Wang, and Weiqiang Yang.
\newblock {Cosmic constraint on massive neutrinos in viable $f(R)$ gravity with
  producing $\Lambda$CDM background expansion}.
\newblock {\em Eur. Phys. J. C}, 76(12):679, 2016.

\bibitem{Hada:2016dje}
Ryuichiro Hada and Toshifumi Futamase.
\newblock {Constraints on neutrino masses from the lensing dispersion of Type
  Ia supernovae}.
\newblock {\em Astrophys. J.}, 828(2):112, 2016.

\bibitem{Wang:2016tsz}
Sai Wang, Yi-Fan Wang, Dong-Mei Xia, and Xin Zhang.
\newblock {Impacts of dark energy on weighing neutrinos: mass hierarchies
  considered}.
\newblock {\em Phys. Rev. D}, 94(8):083519, 2016.

\bibitem{Kumar:2016zpg}
Suresh Kumar and Rafael~C. Nunes.
\newblock {Probing the interaction between dark matter and dark energy in the
  presence of massive neutrinos}.
\newblock {\em Phys. Rev. D}, 94(12):123511, 2016.

\bibitem{Zhao:2016ecj}
Ming-Ming Zhao, Yun-He Li, Jing-Fei Zhang, and Xin Zhang.
\newblock {Constraining neutrino mass and extra relativistic degrees of freedom
  in dynamical dark energy models using Planck 2015 data in combination with
  low-redshift cosmological probes: basic extensions to \ensuremath{\Lambda}CDM
  cosmology}.
\newblock {\em Mon. Not. Roy. Astron. Soc.}, 469(2):1713--1724, 2017.

\bibitem{Bohringer:2016fcq}
Hans B\"ohringer and Gayoung Chon.
\newblock {Constraints on neutrino masses from the study of the nearby
  large-scale structure and galaxy cluster counts}.
\newblock {\em Mod. Phys. Lett. A}, 31(21):1640008, 2016.

\bibitem{Xu:2016ddc}
Lixin Xu and Qing-Guo Huang.
\newblock {Detecting the Neutrinos Mass Hierarchy from Cosmological Data}.
\newblock {\em Sci. China Phys. Mech. Astron.}, 61(3):039521, 2018.

\bibitem{Vagnozzi:2017ovm}
Sunny Vagnozzi, Elena Giusarma, Olga Mena, Katherine Freese, Martina Gerbino,
  Shirley Ho, and Massimiliano Lattanzi.
\newblock {Unveiling $\nu$ secrets with cosmological data: neutrino masses and
  mass hierarchy}.
\newblock {\em Phys. Rev. D}, 96(12):123503, 2017.

\bibitem{Guo:2017hea}
Rui-Yun Guo, Yun-He Li, Jing-Fei Zhang, and Xin Zhang.
\newblock {Weighing neutrinos in the scenario of vacuum energy interacting with
  cold dark matter: application of the parameterized post-Friedmann approach}.
\newblock {\em JCAP}, 05:040, 2017.

\bibitem{Zhang:2017rbg}
Xin Zhang.
\newblock {Weighing neutrinos in dynamical dark energy models}.
\newblock {\em Sci. China Phys. Mech. Astron.}, 60(6):060431, 2017.

\bibitem{Chen:2017ayg}
Lu~Chen, Qing-Guo Huang, and Ke~Wang.
\newblock {New cosmological constraints with extended-Baryon Oscillation
  Spectroscopic Survey DR14 quasar sample}.
\newblock {\em Eur. Phys. J. C}, 77(11):762, 2017.

\bibitem{Yang:2017amu}
Weiqiang Yang, Rafael~C. Nunes, Supriya Pan, and David~F. Mota.
\newblock {Effects of neutrino mass hierarchies on dynamical dark energy
  models}.
\newblock {\em Phys. Rev. D}, 95(10):103522, 2017.

\bibitem{Koksbang:2017rux}
S.~M. Koksbang and S.~Hannestad.
\newblock {Constraining dynamical neutrino mass generation with cosmological
  data}.
\newblock {\em JCAP}, 09:014, 2017.

\bibitem{Li:2017iur}
En-Kun Li, Hongchao Zhang, Minghui Du, Zhi-Huan Zhou, and Lixin Xu.
\newblock {Probing the Neutrino Mass Hierarchy beyond $\Lambda$CDM Model}.
\newblock {\em JCAP}, 08:042, 2018.

\bibitem{Wang:2017htc}
Sai Wang, Yi-Fan Wang, and Dong-Mei Xia.
\newblock {Constraints on the sum of neutrino masses using cosmological data
  including the latest extended Baryon Oscillation Spectroscopic Survey DR14
  quasar sample}.
\newblock {\em Chin. Phys. C}, 42(6):065103, 2018.

\bibitem{Zhao:2017jma}
Ming-Ming Zhao, Jing-Fei Zhang, and Xin Zhang.
\newblock {Measuring growth index in a universe with massive neutrinos: A
  revisit of the general relativity test with the latest observations}.
\newblock {\em Phys. Lett. B}, 779:473--478, 2018.

\bibitem{Boyle:2017lzt}
Aoife Boyle and Eiichiro Komatsu.
\newblock {Deconstructing the neutrino mass constraint from galaxy redshift
  surveys}.
\newblock {\em JCAP}, 03:035, 2018.

\bibitem{Vagnozzi:2018jhn}
Sunny Vagnozzi, Suhail Dhawan, Martina Gerbino, Katherine Freese, Ariel Goobar,
  and Olga Mena.
\newblock {Constraints on the sum of the neutrino masses in dynamical dark
  energy models with $w(z) \geq -1$ are tighter than those obtained in
  $\Lambda$CDM}.
\newblock {\em Phys. Rev. D}, 98(8):083501, 2018.

\bibitem{Guo:2018gyo}
Rui-Yun Guo, Jing-Fei Zhang, and Xin Zhang.
\newblock {Exploring neutrino mass and mass hierarchy in the scenario of vacuum
  energy interacting with cold dark matte}.
\newblock {\em Chin. Phys. C}, 42(9):095103, 2018.

\bibitem{RoyChoudhury:2018gay}
Shouvik Roy~Choudhury and Sandhya Choubey.
\newblock {Updated Bounds on Sum of Neutrino Masses in Various Cosmological
  Scenarios}.
\newblock {\em JCAP}, 09:017, 2018.

\bibitem{Feng:2019mym}
Lu~Feng, Hai-Li Li, Jing-Fei Zhang, and Xin Zhang.
\newblock {Exploring neutrino mass and mass hierarchy in interacting dark
  energy models}.
\newblock {\em Sci. China Phys. Mech. Astron.}, 63(2):220401, 2020.

\bibitem{Zhang:2019ipd}
Jing-Fei Zhang, Bo~Wang, and Xin Zhang.
\newblock {Forecast for weighing neutrinos in cosmology with SKA}.
\newblock {\em Sci. China Phys. Mech. Astron.}, 63(8):280411, 2020.

\bibitem{Li:2020gtk}
Hai-Li Li, Jing-Fei Zhang, and Xin Zhang.
\newblock {Constraints on neutrino mass in the scenario of vacuum energy
  interacting with cold dark matter after Planck 2018}.
\newblock {\em Commun. Theor. Phys.}, 72(12):125401, 2020.

\bibitem{Zhang:2020mox}
Ming Zhang, Jing-Fei Zhang, and Xin Zhang.
\newblock {Impacts of dark energy on constraining neutrino mass after Planck
  2018}.
\newblock {\em Commun. Theor. Phys.}, 72(12):125402, 2020.

\bibitem{deHolanda:2010am}
P.~C. de~Holanda and A.~Yu. Smirnov.
\newblock {Solar neutrino spectrum, sterile neutrinos and additional radiation
  in the Universe}.
\newblock {\em Phys. Rev. D}, 83:113011, 2011.

\bibitem{Palazzo:2013me}
Antonio Palazzo.
\newblock {Phenomenology of light sterile neutrinos: a brief review}.
\newblock {\em Mod. Phys. Lett. A}, 28:1330004, 2013.

\bibitem{Hamann:2013iba}
Jan Hamann and Jasper Hasenkamp.
\newblock {A new life for sterile neutrinos: resolving inconsistencies using
  hot dark matter}.
\newblock {\em JCAP}, 10:044, 2013.

\bibitem{Wyman:2013lza}
Mark Wyman, Douglas~H. Rudd, R.~Ali Vanderveld, and Wayne Hu.
\newblock {Neutrinos Help Reconcile Planck Measurements with the Local
  Universe}.
\newblock {\em Phys. Rev. Lett.}, 112(5):051302, 2014.

\bibitem{Battye:2013xqa}
Richard~A. Battye and Adam Moss.
\newblock {Evidence for Massive Neutrinos from Cosmic Microwave Background and
  Lensing Observations}.
\newblock {\em Phys. Rev. Lett.}, 112(5):051303, 2014.

\bibitem{Dvorkin:2014lea}
Cora Dvorkin, Mark Wyman, Douglas~H. Rudd, and Wayne Hu.
\newblock {Neutrinos help reconcile Planck measurements with both the early and
  local Universe}.
\newblock {\em Phys. Rev. D}, 90(8):083503, 2014.

\bibitem{Archidiacono:2014apa}
Maria Archidiacono, Nicolao Fornengo, Stefano Gariazzo, Carlo Giunti, Steen
  Hannestad, and Marco Laveder.
\newblock {Light sterile neutrinos after BICEP-2}.
\newblock {\em JCAP}, 06:031, 2014.

\bibitem{Ko:2014bka}
P.~Ko and Yong Tang.
\newblock {\ensuremath{\nu}\ensuremath{\Lambda}MDM: A model for sterile
  neutrino and dark matter reconciles cosmological and neutrino oscillation
  data after BICEP2}.
\newblock {\em Phys. Lett. B}, 739:62--67, 2014.

\bibitem{Li:2014dja}
Yun-He Li, Jing-Fei Zhang, and Xin Zhang.
\newblock {Tilt of primordial gravitational wave spectrum in a universe with
  sterile neutrinos}.
\newblock {\em Sci. China Phys. Mech. Astron.}, 57:1455--1459, 2014.

\bibitem{Zhang:2014dxk}
Jing-Fei Zhang, Yun-He Li, and Xin Zhang.
\newblock {Sterile neutrinos help reconcile the observational results of
  primordial gravitational waves from Planck and BICEP2}.
\newblock {\em Phys. Lett. B}, 740:359--363, 2015.

\bibitem{Archidiacono:2014nda}
Maria Archidiacono, Steen Hannestad, Rasmus~Sloth Hansen, and Thomas Tram.
\newblock {Cosmology with self-interacting sterile neutrinos and dark matter -
  A pseudoscalar model}.
\newblock {\em Phys. Rev. D}, 91(6):065021, 2015.

\bibitem{Bergstrom:2014fqa}
Johannes Bergstr\"om, M.~C. Gonzalez-Garcia, V.~Niro, and J.~Salvado.
\newblock {Statistical tests of sterile neutrinos using cosmology and
  short-baseline data}.
\newblock {\em JHEP}, 10:104, 2014.

\bibitem{DayaBay:2014fct}
F.~P. An et~al.
\newblock {Search for a Light Sterile Neutrino at Daya Bay}.
\newblock {\em Phys. Rev. Lett.}, 113:141802, 2014.

\bibitem{Zhang:2014ifa}
Jing-Fei Zhang, Jia-Jia Geng, and Xin Zhang.
\newblock {Neutrinos and dark energy after Planck and BICEP2: data consistency
  tests and cosmological parameter constraints}.
\newblock {\em JCAP}, 10:044, 2014.

\bibitem{Zhang:2014lfa}
Jing-Fei Zhang, Yun-He Li, and Xin Zhang.
\newblock {Measuring growth index in a universe with sterile neutrinos}.
\newblock {\em Phys. Lett. B}, 739:102--105, 2014.

\bibitem{Li:2015poa}
Yun-He Li, Jing-Fei Zhang, and Xin Zhang.
\newblock {Probing $f(R)$ cosmology with sterile neutrinos via measurements of
  scale-dependent growth rate of structure}.
\newblock {\em Phys. Lett. B}, 744:213--217, 2015.

\bibitem{Feng:2017nss}
Lu~Feng, Jing-Fei Zhang, and Xin Zhang.
\newblock {A search for sterile neutrinos with the latest cosmological
  observations}.
\newblock {\em Eur. Phys. J. C}, 77(6):418, 2017.

\bibitem{Zhao:2017urm}
Ming-Ming Zhao, Dong-Ze He, Jing-Fei Zhang, and Xin Zhang.
\newblock {Search for sterile neutrinos in holographic dark energy cosmology:
  Reconciling Planck observation with the local measurement of the Hubble
  constant}.
\newblock {\em Phys. Rev. D}, 96(4):043520, 2017.

\bibitem{Feng:2017mfs}
Lu~Feng, Jing-Fei Zhang, and Xin Zhang.
\newblock {Searching for sterile neutrinos in dynamical dark energy
  cosmologies}.
\newblock {\em Sci. China Phys. Mech. Astron.}, 61(5):050411, 2018.

\bibitem{Feng:2017usu}
Lu~Feng, Jing-Fei Zhang, and Xin Zhang.
\newblock {Search for sterile neutrinos in a universe of vacuum energy
  interacting with cold dark matter}.
\newblock {\em Phys. Dark Univ.}, 23:100261, 2019.

\bibitem{Knee:2018rvj}
Alan~M. Knee, Dagoberto Contreras, and Douglas Scott.
\newblock {Cosmological constraints on sterile neutrino oscillations from
  Planck}.
\newblock {\em JCAP}, 07:039, 2019.

\bibitem{Feng:2019jqa}
Lu~Feng, Dong-Ze He, Hai-Li Li, Jing-Fei Zhang, and Xin Zhang.
\newblock {Constraints on active and sterile neutrinos in an interacting dark
  energy cosmology}.
\newblock {\em Sci. China Phys. Mech. Astron.}, 63(9):290404, 2020.

\bibitem{Feng:2021ipq}
Lu~Feng, Rui-Yun Guo, Jing-Fei Zhang, and Xin Zhang.
\newblock {Cosmological search for sterile neutrinos after Planck 2018}.
\newblock {\em Phys. Lett. B}, 827:136940, 2022.

\bibitem{DiValentino:2021rjj}
Eleonora Di~Valentino, Stefano Gariazzo, Carlo Giunti, Olga Mena, Supriya Pan,
  and Weiqiang Yang.
\newblock {Minimal dark energy: Key to sterile neutrino and Hubble constant
  tensions?}
\newblock {\em Phys. Rev. D}, 105(10):103511, 2022.

\bibitem{Chernikov:2022mdn}
P.~A. Chernikov and A.~V. Ivanchik.
\newblock {The Influence of the Effective Number of Active and Sterile
  Neutrinos on the Determination of the Values of Cosmological Parameters}.
\newblock {\em Astron. Lett.}, 48(12):689--701, 2022.

\bibitem{Pan:2023frx}
Supriya Pan, Osamu Seto, Tomo Takahashi, and Yo~Toda.
\newblock {Constraints on sterile neutrinos and the cosmological tensions}.
\newblock 12 2023.

\bibitem{Verde:2019ivm}
L.~Verde, T.~Treu, and A.~G. Riess.
\newblock {Tensions between the Early and the Late Universe}.
\newblock {\em Nature Astron.}, 3:891, 7 2019.

\bibitem{Vitale:2018yhm}
Salvatore Vitale, Will~M. Farr, Ken Ng, and Carl~L. Rodriguez.
\newblock {Measuring the star formation rate with gravitational waves from
  binary black holes}.
\newblock {\em Astrophys. J. Lett.}, 886(1):L1, 2019.

\bibitem{Yang:2021qge}
Tao Yang.
\newblock {Gravitational-Wave Detector Networks: Standard Sirens on Cosmology
  and Modified Gravity Theory}.
\newblock {\em JCAP}, 05:044, 2021.

\bibitem{Belgacem:2019tbw}
Enis Belgacem, Yves Dirian, Stefano Foffa, Eric~J. Howell, Michele Maggiore,
  and Tania Regimbau.
\newblock {Cosmology and dark energy from joint gravitational wave-GRB
  observations}.
\newblock {\em JCAP}, 08:015, 2019.

\bibitem{Chen:2018rzo}
Zu-Cheng Chen, Fan Huang, and Qing-Guo Huang.
\newblock {Stochastic Gravitational-wave Background from Binary Black Holes and
  Binary Neutron Stars and Implications for LISA}.
\newblock {\em Astrophys. J.}, 871(1):97, 2019.

\bibitem{Du:2021fmb}
Minghui Du and Lixin Xu.
\newblock {How will our knowledge of short gamma-ray bursts affect the distance
  measurement of binary neutron stars?}
\newblock {\em Sci. China Phys. Mech. Astron.}, 65(1):219811, 2022.

\bibitem{deSouza:2019ype}
Josiel Mendon\c{c}a~Soares de~Souza and Riccardo Sturani.
\newblock {Cosmological model selection from standard siren detections by
  third-generation gravitational wave observatories}.
\newblock {\em Phys. Dark Univ.}, 32:100830, 2021.

\bibitem{Regimbau:2016ike}
T.~Regimbau, M.~Evans, N.~Christensen, E.~Katsavounidis, B.~Sathyaprakash, and
  S.~Vitale.
\newblock {Digging deeper: Observing primordial gravitational waves below the
  binary black hole produced stochastic background}.
\newblock {\em Phys. Rev. Lett.}, 118(15):151105, 2017.

\bibitem{Belgacem:2018lbp}
Enis Belgacem, Yves Dirian, Stefano Foffa, and Michele Maggiore.
\newblock {Modified gravitational-wave propagation and standard sirens}.
\newblock {\em Phys. Rev. D}, 98(2):023510, 2018.

\bibitem{Safarzadeh:2019pis}
Mohammadtaher Safarzadeh, Edo Berger, Ken K.~Y. Ng, Hsin-Yu Chen, Salvatore
  Vitale, Chris Whittle, and Evan Scannapieco.
\newblock {Measuring the delay time distribution of binary neutron stars. II.
  Using the redshift distribution from third-generation gravitational wave
  detectors network}.
\newblock {\em Astrophys. J. Lett.}, 878(1):L13, 2019.

\bibitem{Song:2019ddw}
Hao-Ran Song, Shun-Ke Ai, Min-Hao Wang, Nan Xing, He~Gao, and Bing Zhang.
\newblock {Viewing angle constraints on S190425z and S190426c and the joint
  gravitational-wave/gamma-ray detection fractions for binary neutron star
  mergers}.
\newblock {\em Astrophys. J. Lett.}, 881(2):L40, 2019.

\bibitem{Wanderman:2014eza}
David Wanderman and Tsvi Piran.
\newblock {The rate, luminosity function and time delay of non-Collapsar short
  GRBs}.
\newblock {\em Mon. Not. Roy. Astron. Soc.}, 448(4):3026--3037, 2015.

\bibitem{Yu:2021nvx}
Jiming Yu, Haoran Song, Shunke Ai, He~Gao, Fayin Wang, Yu~Wang, Youjun Lu,
  Wenjuan Fang, and Wen Zhao.
\newblock {Multimessenger Detection Rates and Distributions of Binary Neutron
  Star Mergers and Their Cosmological Implications}.
\newblock {\em Astrophys. J.}, 916(1):54, 2021.

\bibitem{Regimbau:2014nxa}
T.~Regimbau, K.~Siellez, D.~Meacher, B.~Gendre, and M.~Bo\"er.
\newblock {Revisiting coincidence rate between Gravitational Wave detection and
  short Gamma-Ray Burst for the Advanced and third generation}.
\newblock {\em Astrophys. J.}, 799(1):69, 2015.

\bibitem{Madau:2014bja}
Piero Madau and Mark Dickinson.
\newblock {Cosmic Star Formation History}.
\newblock {\em Ann. Rev. Astron. Astrophys.}, 52:415--486, 2014.

\bibitem{Eichhorn:2018phj}
Astrid Eichhorn, Tim Koslowski, and Antonio~D. Pereira.
\newblock {Status of background-independent coarse-graining in tensor models
  for quantum gravity}.
\newblock {\em Universe}, 5(2):53, 2019.

\bibitem{KAGRA:2021duu}
R.~Abbott et~al.
\newblock {Population of Merging Compact Binaries Inferred Using Gravitational
  Waves through GWTC-3}.
\newblock {\em Phys. Rev. X}, 13(1):011048, 2023.

\bibitem{LIGOScientific:2018mvr}
B.~P. Abbott et~al.
\newblock {GWTC-1: A Gravitational-Wave Transient Catalog of Compact Binary
  Mergers Observed by LIGO and Virgo during the First and Second Observing
  Runs}.
\newblock {\em Phys. Rev. X}, 9(3):031040, 2019.

\bibitem{Ozel:2016oaf}
Feryal \"Ozel and Paulo Freire.
\newblock {Masses, Radii, and the Equation of State of Neutron Stars}.
\newblock {\em Ann. Rev. Astron. Astrophys.}, 54:401--440, 2016.

\bibitem{Zhang:2017srh}
Xing Zhang, Tan Liu, and Wen Zhao.
\newblock {Gravitational radiation from compact binary systems in screened
  modified gravity}.
\newblock {\em Phys. Rev. D}, 95(10):104027, 2017.

\bibitem{Zhao:2017cbb}
Wen Zhao and Linqing Wen.
\newblock {Localization accuracy of compact binary coalescences detected by the
  third-generation gravitational-wave detectors and implication for cosmology}.
\newblock {\em Phys. Rev. D}, 97(6):064031, 2018.

\bibitem{Wen:2010cr}
Linqing Wen and Yanbei Chen.
\newblock {Geometrical Expression for the Angular Resolution of a Network of
  Gravitational-Wave Detectors}.
\newblock {\em Phys. Rev. D}, 81:082001, 2010.

\bibitem{Cutler:1992tc}
Curt Cutler et~al.
\newblock {The Last three minutes: issues in gravitational wave measurements of
  coalescing compact binaries}.
\newblock {\em Phys. Rev. Lett.}, 70:2984--2987, 1993.

\bibitem{Sathyaprakash:2009xs}
B.~S. Sathyaprakash and B.~F. Schutz.
\newblock {Physics, Astrophysics and Cosmology with Gravitational Waves}.
\newblock {\em Living Rev. Rel.}, 12:2, 2009.

\bibitem{Howell:2018nhu}
E.~J. Howell, K.~Ackley, A.~Rowlinson, and D.~Coward.
\newblock {Joint gravitational wave -- gamma-ray burst detection rates in the
  aftermath of GW170817}.
\newblock 11 2018.

\bibitem{Tan:2020vtc}
Wei-Wei Tan and Yun-Wei Yu.
\newblock {The jet structure and the intrinsic luminosity function of short
  gamma-ray bursts}.
\newblock {\em Astrophys. J.}, 902(1):83, 2020.

\bibitem{Stratta:2018ldl}
G.~Stratta, L.~Amati, R.~Ciolfi, and S.~Vinciguerra.
\newblock {THESEUS in the era of Multi-Messenger Astronomy}.
\newblock {\em Mem. Soc. Ast. It.}, 89(2):205--212, 2018.

\bibitem{Speri:2020hwc}
Lorenzo Speri, Nicola Tamanini, Robert~R. Caldwell, Jonathan~R. Gair, and
  Benjamin Wang.
\newblock {Testing the Quasar Hubble Diagram with LISA Standard Sirens}.
\newblock {\em Phys. Rev. D}, 103(8):083526, 2021.

\bibitem{Hirata:2010ba}
Christopher~M. Hirata, Daniel~E. Holz, and Curt Cutler.
\newblock {Reducing the weak lensing noise for the gravitational wave Hubble
  diagram using the non-Gaussianity of the magnification distribution}.
\newblock {\em Phys. Rev. D}, 81:124046, 2010.

\bibitem{Kocsis:2005vv}
Bence Kocsis, Zsolt Frei, Zoltan Haiman, and Kristen Menou.
\newblock {Finding the electromagnetic counterparts of cosmological standard
  sirens}.
\newblock {\em Astrophys. J.}, 637:27--37, 2006.

\bibitem{Planck:2018vyg}
N.~Aghanim et~al.
\newblock {Planck 2018 results. VI. Cosmological parameters}.
\newblock {\em Astron. Astrophys.}, 641:A6, 2020.
\newblock [Erratum: Astron.Astrophys. 652, C4 (2021)].

\bibitem{Beutler:2011hx}
Florian Beutler, Chris Blake, Matthew Colless, D.~Heath Jones, Lister
  Staveley-Smith, Lachlan Campbell, Quentin Parker, Will Saunders, and Fred
  Watson.
\newblock {The 6dF Galaxy Survey: Baryon Acoustic Oscillations and the Local
  Hubble Constant}.
\newblock {\em Mon. Not. Roy. Astron. Soc.}, 416:3017--3032, 2011.

\bibitem{Ross:2014qpa}
Ashley~J. Ross, Lado Samushia, Cullan Howlett, Will~J. Percival, Angela Burden,
  and Marc Manera.
\newblock {The clustering of the SDSS DR7 main Galaxy sample \textendash{} I. A
  4 per cent distance measure at $z = 0.15$}.
\newblock {\em Mon. Not. Roy. Astron. Soc.}, 449(1):835--847, 2015.

\bibitem{BOSS:2016wmc}
Shadab Alam et~al.
\newblock {The clustering of galaxies in the completed SDSS-III Baryon
  Oscillation Spectroscopic Survey: cosmological analysis of the DR12 galaxy
  sample}.
\newblock {\em Mon. Not. Roy. Astron. Soc.}, 470(3):2617--2652, 2017.

\bibitem{Pan-STARRS1:2017jku}
D.~M. Scolnic et~al.
\newblock {The Complete Light-curve Sample of Spectroscopically Confirmed SNe
  Ia from Pan-STARRS1 and Cosmological Constraints from the Combined Pantheon
  Sample}.
\newblock {\em Astrophys. J.}, 859(2):101, 2018.

\bibitem{Lewis:2002ah}
Antony Lewis and Sarah Bridle.
\newblock {Cosmological parameters from CMB and other data: A Monte Carlo
  approach}.
\newblock {\em Phys. Rev. D}, 66:103511, 2002.

\bibitem{ETcurve-web}
\url{https://www.et-gw.eu/index.php/etsensitivities/}.

\bibitem{CEcurve-web}
\url{https://cosmicexplorer.org/sensitivity.html}.

\bibitem{Zhu:2021ram}
Jin-Ping Zhu et~al.
\newblock {Kilonovae and Optical Afterglows from Binary Neutron Star Mergers.
  II. Optimal Search Strategy for Serendipitous Observations and
  Target-of-opportunity Observations of Gravitational Wave Triggers}.
\newblock {\em Astrophys. J.}, 942(2):88, 2023.

\end{thebibliography}

\end{document}